\documentclass[final,3p,times]{elsarticle}
\usepackage{amssymb}
\usepackage{amsmath}
\usepackage{orcidlink}
\usepackage{soul}
\usepackage{algorithm}
\usepackage{algpseudocode}
\usepackage{comment} 
\usepackage{graphicx}   % incluir imágenes
\usepackage{float}      % para [htp!]
\usepackage{caption}       % Para subtítulos y captions avanzados
\usepackage[table]{xcolor} % Para colorear celdas y filas de tablas
\usepackage{array}         % Para controlar alineamiento y anchura de columnas en tablas
\usepackage{colortbl}      % Para usar colores dentro de tablas (opcional, complementa xcolor) 

\usepackage{stfloats}   % o \usepackage{dblfloatfix}
\usepackage{booktabs} % poner en el preámbulo

\usepackage{tikz}       % dibujo vectorial
\usetikzlibrary{positioning} % útil si quieres
\usepackage{subcaption}
\usetikzlibrary{positioning}
\usetikzlibrary{fit}
\usetikzlibrary{arrows.meta}
\usepackage{tabularx}
\usepackage{multicol}
\usepackage[capitalise,noabbrev]{cleveref}
\usepackage{soul}
\usepackage[export]{adjustbox}
\usepackage{lineno}
\usetikzlibrary{external}
\journal{\href{https://www.sciencedirect.com/journal/acta-astronautica}{Acta Astronautica}}

\begin{document}

\begin{frontmatter}



\title{International Space Station operational modal analysis via iterative pole relocation}


\author[label1]{Marco Civera \orcidlink{0000-0003-0414-7440}\corref{cor2}\fnref{label4}}
\author[label2]{Gabriele Dessena \orcidlink{0000-0001-7394-9303}\corref{cor1}\fnref{label4}}

\author[label2]{Marina {Cózar Alcázar}} % family names/Surnames: Cózar Alcázar

\author[label2]{Saray {Undiano Echániz}} % family names/Surnames: Undiano Echániz

\affiliation[label1]{organization={Department of Structural, Building and Geotechnical Engineering, Politecnico di Torino},%Department and Organization
            addressline={Corso Duca degli Abruzzi 24}, 
            city={Turin},
            postcode={10129}, 
            state={Piedmont},
            country={Italy}}
\affiliation[label2]{organization={Department of Aerospace Engineering, Universidad Carlos III de Madrid},
             addressline={Av.da de la Universidad 30},
             postcode={28911},
             city={Leganés},
             state={Madrid},
             country={Spain}}

\author[label3]{Oscar E. Bonilla-Manrique \orcidlink{0000-0003-0541-8310}}
\affiliation[label3]{organization={Electronic Technology Department, Universidad Carlos III de Madrid},
             addressline={Av.da de la Universidad 30},
             postcode={28911},
             city={Leganés},
             state={Madrid},
             country={Spain}}

\fntext[label4]{These authors contributed equally to this work.}
\cortext[cor1]{Corresponding author: \href{mailto:gdessena@ing.uc3m.es}{gdessena@ing.uc3m.es}}
%% Abstract
\begin{abstract}
In recent years, increasing aerospace safety requirements have intensified the demand for reliable structural damage detection. This work presents an Operational Modal Analysis approach for accurate modal parameter estimation, with an application to space structure monitoring. The proposed System Identification (SI) method innovatively combines the Natural Excitation Technique (NExT) with the Fast and Relaxed Vector Fitting (FRVF) algorithm, which uses an iterative least-squares optimisation. A preliminary validation is first carried out on a numerical beam model, comparing results with analytical solutions and the established Natural Excitation Technique with Eigensystem Realisation Algorithm (NExT-ERA) and Stochastic Subspace Identification with Canonical Variate Analysis (SSI) methods. Then, operational validation is performed on real acceleration data from the Space Acceleration Measurement Systems aboard the International Space Station. Identified vibration modes from NExT-FRVF and NExT-ERA show comparable results after signal processing, with mode consistency assessed by repeated occurrence and physical interpretation, while SSI fails to identify most. The output-only algorithm proves to be highly reliable, outperforming benchmark methods under noisy conditions on a numerical system and offering reliable identifications on the experimental data.
\end{abstract}



\begin{highlights}
\item This work introduces a novel FRVF-based output-only identification algorithm.
\item Optimised configuration improves accuracy with respect to NExT-ERA.
\item Robust mode identification under different noise conditions.
\item Experimental validation on the International Space Station.
\item The extracted modes are shown to be consistent and physically meaningful.
\end{highlights}

%% Keywords
\begin{keyword}
 International Space Station \sep System Identification \sep Structural Health Monitoring \sep Operational Modal Analysis \sep Output-Only \sep Modal Parameters \sep Natural Excitation Technique \sep Fast and Relaxed Vector Fitting



\end{keyword}

\end{frontmatter}

%\linenumbers

\section{Introduction}

Space structure environments are classified mainly as ground, launch, and in-orbit loads. Throughout these phases, an accurate understanding of the dynamic characteristics of a target system is crucial to assess its structural conditions, integrity, and performance. This knowledge is typically obtained through System Identification (SI), which is typically used to estimate structural parameters from experimental data \cite{Chakraborty2022}. Such data can, for instance, support pointing control and monitoring tasks. The latter is the focus of Structural Health Monitoring (SHM), where this information is used to detect, localise, and evaluate potential damage throughout the operational life of the structure. Thus, achieving unequivocal SHM for space structures requires efficient and accurate SI methods.

The capability to properly capture and control the dynamics of the coupled system requires precise identification of its mass and structural properties \cite{A.J.Elliott2025}. Traditionally, these modal characteristics are obtained through ground tests; however, they can vary significantly under operational conditions \cite{Sabatini2013}. Therefore, implementing Operational Modal Analysis (OMA) can be beneficial to track these parameters throughout the lifetime of the mission.

These approaches aim to ensure structural integrity and prevent performance degradation in modern spacecraft, including large, lightweight, and flexible structures such as antennas and solar arrays, such as for the International Space Station (ISS) \cite{Djojodihardjo2024}. The dynamic behaviour of such structures is further complicated by orbit-attitude-vibration coupling effects, as demonstrated analytically for symmetric and asymmetric configurations of China's Tiangong (Chinese Space Station) \cite{Yang2026}, which reinforces the need for accurate in-orbit modal identification to support both attitude control and structural integrity assessment. Consequently, SHM represents both a challenging and promising research area for the next generation of aerospace systems \cite{Iannelli2022}.

\subsection{System identification: Experimental and operational modal analysis}

Depending on whether excitation signals are known or not, SI methods can be input-output or output-only. These can be used, respectively, for Experimental Modal Analysis (EMA) and Operational Modal Analysis (OMA). EMA assumptions mandate a known experimental setup, where the dynamic input is known and, possibly, controlled (e.g., by means of shaker or hammer tests). Instead, this work deals with operational scenarios, so EMA is not discussed further. The interested reader is referred to \cite{Avitabile2017} for the necessary background on classic EMA procedures.

In OMA, the modal parameters are estimated using only the structural response. This allows identification of systems where the application of controlled input is impractical due to their size or operational criteria, such as bridges \cite{Magalhaes2011}, buildings \cite{Dessena2022}, and, as in this case, aerospace vehicles \cite{Sibille2023}.  In all these applications, the structures are excited by ambient forces, such as wind or traffic, which cannot be directly measured, nor isolated or neglected. OMA typically assumes: (i) linearity, (ii) stationarity of system properties, and (iii) observability, which, in practical terms, means sufficient sensor coverage. These assumptions can generally be satisfied for many space structures because (i) the vibration amplitude is small (during most of its in-orbit life), (ii) a suitable, short or long, time window can approximate stationarity even in time-varying systems, and (iii) the positioning of the sensor is known and permanent.

Typically, ambient excitation is modelled as White Gaussian Noise (WGN) \cite{Reynders2012}. This enables the use of techniques such as the Natural Excitation Technique (NExT)  \cite{GeorgeIII1993} (discussed in detail later) to generate correlation functions to extract an Impulse Response Function (IRF). Alternatively, stochastic methods, such as Stochastic Subspace Identification with Canonical Variate Analysis (SSI), can be used. Furthermore, methods arising from other domains have also been extended to OMA to exploit their computational performance and noise robustness capabilities. An example is the Loewner Framework, first introduced as a frequency-domain reduced order modelling technique in electrical engineering \cite{Lefteriu2010b}, and then extended to OMA in reinforced concrete buildings \cite{Dessena2024f} and helicopter blades \cite{Dessena2026}. Another notable recently extended method is Dynamic Mode Decomposition (DMD). First introduced for aerodynamic flow modelling in \cite{Schmid2010}, DMD is extended to modal analysis in \cite{Saito2020}, and applied to EMA on an aircraft model in \cite{Wu2025} and, finally, to OMA applications on aeronautical structures in \cite{Wu2026}.

\subsection{Operational modal analysis for modal tracking and damage assessment}
OMA is the enabling technology behind modal tracking and, in turn, modern vibration-based SHM \cite{Farrar2025}, which generally uses modal parameters as an indicator of the presence, intensity, and location of damage. In aerospace engineering, SI techniques have been employed for SHM purposes in various structures. Most notably, they have been applied for the modal identification of the Space Station Resource Node to ensure parameter consistency \cite{Pappa1999}, and for the Space Shuttle tail rudder to enable continuous vibration-based SHM monitoring \cite{Pappa1998}. 

\begin{figure}[htp!]
\centering
{\includegraphics[width=\textwidth]{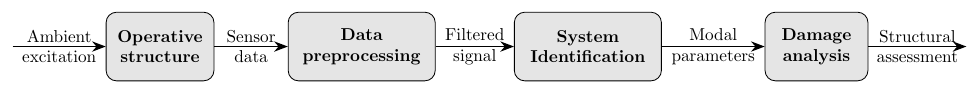}}
\caption{Output-only SI and vibration-based SHM process.}
\label{f11}
\end{figure}

Under varying conditions, the parameters identified through OMA are used to detect changes that can indicate damage, assess severity, estimate remaining life, and optimise maintenance.  This is possible since key modal parameters -- natural frequencies ($\omega_n$), damping ratios ($\zeta_n$), and mode shapes ($\mathbf{\phi}_n$) -- are directly related to the mass ($\mathbf{M}$), stiffness ($\mathbf{K}$), and damping ($\mathbf{C}_d$) matrices \cite{Civera2021a}. Frequency shifts often indicate damage \cite{Hong-ping2005}, while $\mathbf{\phi}_n$ can support its localisation \cite{Porcu2019}. Combining both improves diagnosis, though damping ratios are less reliable in output-only contexts due to noise sensitivity. Consequently, SHM can be considered as a non-destructive evaluation of a system \cite{Balageas2006}. The general process that links both disciplines, SI and SHM, is shown in Figure \ref{f11}.

\subsection{Aim and objectives of this work}
Given the continuing and everlasting need for computationally robust and efficient output-only SI methods and the potential and significant added value that vibration-based SHM can provide in space and on-orbit applications, this work proposes a novel OMA algorithm for output-only SI, specifically tested and validated to operate under such extreme environmental conditions.

The proposed OMA method is based on the extension of the Fast and Relaxed Vector Fitting (FRVF) from Single-Input Multi-Output (SIMO) modal analysis in the frequency domain to OMA, using NExT. FRVF originates from electrical engineering, where it is used for the modelling of electrical circuits \cite{Gustavsen2006a} and has received wide recognition and many applications and adaptations over the years. Thanks to mechanical–electrical analogies, it has been extended to the extraction of modal parameters from EMA, with the earliest applications to metal frames \cite{Civera2021} and masonry bridges \cite{Civera2021a}. 
Already applied to SIMO EMA in aeronautics \cite{Civera2025i}, FRVF shows excellent performance in terms of time-to-identification compared to traditional methods like Numerical algorithms for (4) Subspace state space system IDentification (N4SID), as shown in \cite{Dessena2022g}. 

For benchmarking, this work considers NExT with the Eigensystem Realisation Algorithm (NExT-ERA) \cite{Siringoringo2008} and Stochastic Subspace Identification with Canonical Variate Analysis (SSI) \cite{VanOverschee1996}, the latter on the experimental case study only. NExT-ERA  is widely used in aerospace engineering, e.g., aero-structures \cite{Moncayo2010}, and provides a fair and direct comparison for the proposed NExT-FRVF, while SSI represents an industrial standard time-domain method.  
In this sense, time-domain methods, which rely directly on time histories or correlation functions among them, are generally highly computationally efficient but may be less accurate, as noise and limited data length can reduce the precision of mode estimation. In contrast, frequency-domain methods, such as those based on spectral density functions, can potentially lead to more accurate results.

Despite ISS vibration data having been continuously available since 2001 under the National Aeronautics and Space Administration (NASA) Principal Investigator Microgravity Services (PIMS), to the best of the authors' knowledge, no systematic study has been carried out for its vibration monitoring. Additionally, the existing literature focuses on the modal identification of sub-components, such as its solar arrays \cite{Kerschen2025}, on the dynamic response to specific forced excitation such as engine firing \cite{Fitzpatrick2013}, or on signal characterisation of the accelerometric environment itself, including the detection of nonlinearities in SAMS recordings \cite{Saez2014}. 

For these reasons, this work pioneeringly proposes an output-only implementation of FRVF and the modal identification of the ISS from ambient, in-orbit, excitation, achieving a twofold objective:

\begin{itemize}
    \item \textit{First implementation of output-only FRVF}: The introduction of a novel and efficient OMA method based on the iterative relocation of poles;
    \item \textit{Application to aerospace}: Validation of the proposed NExT-FRVF on the largest inhabited space system, the International Space Station (ISS).
\end{itemize}

The remaining sections of the article are organised as follows:
\begin{enumerate}
    \item[2.] \hyperref[sec:met]{\emph{Methodology}}: FRVF, NExT, and their joint application are introduced;
    \item[3.] \hyperref[sec:num]{\emph{Numerical case study}}: An Euler-Bernoulli cantilever beam, with and without artificially added measurement noise, is used as a preliminary benchmark for NExT-FRVF;
    \item[4.] \hyperref[sec:iss]{\emph{The International Space Station}}: The newly proposed NExT-FRVF method is applied to the identification and monitoring of the ISS;
    \item[5.] The \hyperref[sec:conc]{\emph{Conclusions}} close this paper.
\end{enumerate}

\section{Methodology}\label{sec:met}
In this section, the classic FRVF definition is introduced, before defining NExT and describing how their combination (NExT-FRVF) can be used for OMA. The principles of vibration-based SHM are then briefly recalled as well, for completeness.

\subsection{Fast and Relaxed Vector Fitting}

FRVF is a computationally enhanced version of Vector Fitting (VF). VF fits frequency-domain responses via rational functions. Based on 1960s work on transfer functions (TFs) \cite{Sanathanan1963}, VF was formalised by Gustavsen and Semlyen \cite{Gustavsen1999a} for electrical systems and, later, after its fast and relaxed approach was developed, applied to mechanical and civil structures for modal analysis, as mentioned earlier in the Introduction.

The method is well-known for its robustness in complicated mechanical systems, including multiple resonance peaks, and can be summarised in two main stages \cite{Gustavsen1999a}: Identification of poles and residues using iterative least squares (LS). 
%These are central in the identification of the suitable TF, which can be defined as usual by relating a system FRF, \( H(\omega) \), to its input \( X(\omega) \) and output \( U(\omega) \) spectra, such that:
Let \(u(t)\) and \(y(t)\) denote, respectively, the input and output of a linear time-invariant system. In the frequency domain, their spectra \(U(\omega)\) and \(Y(\omega)\) are related by the transfer function \(H(\omega)\) as:

\begin{equation}
    Y(\omega) = H(\omega) U(\omega)
\end{equation}

Now, for illustration, consider a general single degree-of-freedom (SDOF) mass-spring-damper system, defined as such:

\begin{equation}
    m\ddot{x}(t)+c\dot{x}(t)+kx(t)=F(t),
\end{equation}

where \(x(t)\), \(\dot{x}(t)\), and \(\ddot{x}(t)\) are the displacement, velocity, and acceleration, respectively, \(m\), \(c\), and \(k\) are the mass, damping, and stiffness coefficients, in the same order, and \(F(t)\) is the external driving force. 

Assuming zero initial conditions, the displacement-to-force transfer function (i.e., the Receptance FRF) is
\begin{equation}
H_x(s)=\frac{X(s)}{F(s)}
      =\frac{1}{ms^2+cs+k}
      =\frac{1}{m(s-s_1)(s-s_2)},
\end{equation}
where
\begin{equation}
s_{1,2}=-\zeta\omega_n \pm \mathrm{i}\omega_d,
\qquad
\omega_d=\omega_n\sqrt{1-\zeta^2}.
\end{equation}
and the corresponding partial-fraction expansion (PFE) is
\begin{equation}
H_x(s)=\frac{A_1}{s-s_1}+\frac{A_2}{s-s_2},
\end{equation}
with
\begin{equation}
A_1=\frac{1}{m(s_1-s_2)}, \qquad
A_2=\frac{1}{m(s_2-s_1)}.
\end{equation}

Note: If acceleration responses are considered instead of displacement responses, the acceleration-to-force transfer function becomes
\begin{equation}
\ddot{H}(s)=\frac{\ddot{X}(s)}{F(s)}
      =s^2 H_x(s)
      =\frac{s^2}{ms^2+cs+k}.
\end{equation}
also known as the Accelerance FRF. Similarly, output velocities lead to the so-called Mobility FRF \(\dot{H}(s)\). However, in all these cases, the poles are unchanged with respect to \(H_x(s)\), and therefore the natural frequencies and damping ratios are extracted in the same way. However, the residues differ and must be interpreted according to the measured output quantity.)

To fit the experimental FRF, several TF formulations can be considered, all of them with the same core physical meaning but with different characteristics. Although Rational Fraction Polynomial (RFP) and Pole-zero models have been used in the literature for modal identification and structural damage detection \cite{Figueiredo2009}, \cite{Farrar2000}, \cite{Verboven2002}, they are not employed here due to non-convexity \cite{Richardson1985}, multiple local minima, and nonlinearity in their parameters \cite{Civera2024}. Instead, the partial-fraction expansion is adopted. The crucial point is that, for fixed poles, this formulation is linear in the residues, while the poles are updated iteratively by VF:

\begin{equation}
H(s) =
\sum_{n=1}^{N_m}
\left(
\frac{R_n}{s-p_n}
+
\frac{R_n^\ast}{s-p_n^\ast}
\right),
\label{eq222}
\end{equation}

\noindent where \( R_n \) are the identified residues, \( p_n \) are the corresponding poles, and * indicates the complex conjugate. 

In the remainder of the paper, a compact pole-by-pole formulation will be used for this and related formulations, of the kind 

\begin{equation}
H(s) = \sum_{n=1}^{N_p} \frac{R_n}{s-p_n}
\label{eq222b}
\end{equation}

\noindent implying omitting explicit complex conjugation for brevity. That is to say, note that \(p_{n+1}=p_{n}^*\) is the complex conjugate of \(p_{n}\) - poles are identified in pairs. Thus, in the notation used in this paper, the summation goes up to \(N_m\) (the number of conjugate modal pairs, i.e., half the total number of poles \(N_p\)) for Eq. \ref{eq222}, but up to \(N_p\) in Eq. \ref{eq222b} and all the following equations.

\subsubsection{Iterative process mathematical formulation}

For a given set of current (trial) poles \(a_n\), VF introduces a rational numerator function \(h(s)\), written as
\begin{equation}
h(s)=\sum_{n=1}^{N_p}\frac{C_n}{s-a_n}+d+se,
\label{eq223}
\end{equation}
where \(C_n\) are the auxiliary residues associated with the pole-relocation step. The poles \(a_n\) are updated iteratively and, after convergence, approximate the physical poles \(p_n\) of the system. Similarly, the auxiliary residues should converge to the actual residues, i.e., \( C_n \neq R_n \) initially, \( C_n \rightarrow R_n \) iteration after iteration.
However, the final physical residues \(R_n\) can also be recomputed directly after the pole-relocation stage has been completed.

The auxiliary residues \( C_n \) and poles \( a_n \) can be either real or complex conjugate pairs \cite{Gustavsen1999a}. In contrast, the coefficients \( d \) and \( e \), which compensate for out-of-band contributions (modes below or above the investigated frequency band), are real-valued.

A two-stage LS procedure estimates these parameters as follows:

\noindent \textit{Preliminary stage: Initial poles}

\noindent Complex conjugate poles are chosen to avoid convergence problems, with the imaginary part covering the frequency range and the real part small enough $\alpha = \beta/100$ to avoid ill-conditioning:

\begin{equation}
    a_n = -\alpha + \mathrm{i}\beta, \quad a_{n+1} = a_{n}^* =-\alpha - \mathrm{i}\beta
    \label{init}
\end{equation}

\noindent \textit{Stage 1: Pole identification}

\noindent The fitting problem is formulated by introducing an auxiliary scaling function \(\sigma(s)\). Over subsequent iterations, \(\sigma(s)\) tends to one as the residues of the scaling function, denoted by \(\widetilde{C}_n\), tend to zero. Thus, it is important to highlight that the functions \(h(s)\) and \(\sigma(s)\) share the same set of current poles \(a_n\), but not the same set of residues.

Defining \(f(s)\) to denote the measured frequency-response data to be fitted, one has:

\begin{equation}
\begin{cases}
h(s) \approx \sigma(s)f(s), \\[4pt]
h(s)=\displaystyle\sum_{n=1}^{N_p}\frac{C_n}{s-a_n}+d+se, \\[6pt]
\sigma(s)=1+\displaystyle\sum_{n=1}^{N_p}\frac{\widetilde{C}_n}{s-a_n}.
\end{cases}
\label{eq:sistema}
\end{equation}

Rewriting Eq. \ref{eq:sistema} yields a linear and overdetermined LS problem in \( C_n, \widetilde{C}_n, d, e \) over \( N_s \) frequency samples (Strictly speaking, the experimental data are available only at the sampled frequency points \(s_k\); therefore, \(f(s_k)\) denotes the measured value at the \(k\)-th frequency sample.):

\begin{equation}
\sum_{n=1}^{N_p} \frac{C_n}{s_k-a_n}
+d+s_k e
-
f(s_k) \left(1 + \sum_{n=1}^{N_p}\frac{\widetilde{C}_n}{s_k-a_n}\right)
\approx 0,
\qquad k=1,\ldots,N_s .
\label{eq220}
\end{equation}

For a fixed set of current poles \(a_n\), Eq.~\eqref{eq220} is linear in the unknown coefficients \(C_n\), \(\widetilde{C}_n\), \(d\), and \(e\). Then, expressing Eq. \ref{eq220} via partial fractions gives a key form of the scaling function:

\begin{equation}
\sigma(s)=
\frac{\displaystyle\prod_{n=1}^{N_p}(s-z_n)}
     {\displaystyle\prod_{n=1}^{N_p}(s-a_n)},
\end{equation}
where, as said, \(a_n\) are the current poles (solutions of the denominator) and \(z_n\) are the zeros of \(\sigma(s)\) (solutions of the numerator). Since
\begin{equation}
H(s)\approx\frac{h(s)}{\sigma(s)},
\end{equation}
the zeros \(z_n\) of \(\sigma(s)\) become the relocated poles for the next iteration:
\begin{equation}
a_n \leftarrow z_n .
\end{equation}

It follows that the relocated poles are given by the zeros of \(\sigma(s)\). These zeros, \(z_n\), are obtained from the eigenvalue problem:

\begin{equation}
\{z_n\}_{n=1}^{N_p}
=
\mathrm{eig}\left(\mathbf{A}-\mathbf{b}\widetilde{\mathbf{C}}^{T}\right),
\label{eqzeros}
\end{equation}
where
\begin{equation}
\mathbf{A}=\mathrm{diag}(a_1,a_2,\ldots,a_{N_p}),
\qquad
\mathbf{b}=\begin{bmatrix}1&1&\cdots&1\end{bmatrix}^{T},
\end{equation}
and \(\widetilde{\mathbf{C}}^{T}=[\widetilde{C}_1,\widetilde{C}_2,\ldots,\widetilde{C}_{N_p}]\) contains the residues of the scaling function. 

\noindent \textit{Stage 2: Residue identification}

\noindent After the pole-relocation iterations have been completed, the final relocated poles are denoted, as mentioned, by \(p_n\) and kept fixed. Only one pole from each pair is retained when reporting modal parameters. The physical residues \(R_n\), together with \(d\) and \(e\), are finally estimated by fitting the experimental samples:
\begin{equation}
f(s_k)\approx
\sum_{n=1}^{N_p}\frac{R_n}{s_k-p_n}+d+s_k e,
\qquad k=1,\ldots,N_s .
\end{equation}
The final fitted transfer function is therefore
\begin{equation}
H(s)=
\sum_{n=1}^{N_p}\frac{R_n}{s-p_n}+d+se.
\end{equation}

For multi-output data, the mode shape components are inferred from the residue vectors associated with each identified pole. Under the common assumptions of ambient vibration testing, output-only modal analysis determines these mode shapes up to an arbitrary scale factor, but this is not an issue, since the identified mode shapes further need normalisation before comparison. 

 Although the residue amplitudes depend on the input location, excitation level, and response quantity, the relative components across the output channels provide the corresponding mode-shape estimate. Since these estimates are determined up to an arbitrary scale and phase factor, they are normalised before comparison.

More details about this procedure are given in Appendices A and B of \cite{Gustavsen1999a}.

\subsubsection{Relaxed Vector Fitting extension}

The relaxed VF extension aims to mitigate excessively large pole shifts, which may otherwise lead to large values of the scaling function \(\sigma(s)\) and, consequently, to increased least-squares fitting error. In the standard VF formulation, the scaling function is constrained by the asymptotic condition \(\sigma(\infty)=1\). In the relaxed formulation, this condition is replaced by an unknown constant term \(d_\sigma\), giving \cite{Gustavsen2006a}
\begin{equation}
\sigma(s)
=
d_\sigma
+
\sum_{n=1}^{N_p}
\frac{\widetilde{C}_n}{s-a_n},
\label{rex1}
\end{equation}
where, as before, \(\widetilde{C}_n\) are the residues of the scaling function and \(a_n\) are the current trial poles.

With this definition, the pole-identification least-squares problem becomes (compare to the previous Eq. \ref{eq220})

\begin{equation}
\sum_{n=1}^{N_p}
\frac{C_n}{s_k-a_n}
+d+s_k e
-
f(s_k)
\left(
d_\sigma+
\sum_{n=1}^{N_p}
\frac{\widetilde{C}_n}{s_k-a_n}
\right)
\approx 0,
\qquad k=1,\ldots,N_s .
\label{eq:relaxed_ls}
\end{equation}
Since Eq.~\eqref{eq:relaxed_ls} is homogeneous in the unknowns, the trivial solution \(\sigma(s)=0\) must be prevented. This is done by imposing the additional non-triviality constraint
\begin{equation}
\mathrm{Re}
\left\{
\sum_{k=1}^{N_s}
\left(
d_\sigma+
\sum_{n=1}^{N_p}
\frac{\widetilde{C}_n}{s_k-a_n}
\right)
\right\}
=
N_s .
\label{rex2}
\end{equation}

If frequency weighting is used in the least-squares fit, the residual at each frequency sample is multiplied by a scalar weight \(w_k\). The non-triviality constraint in Eq.~\eqref{rex2} is scaled consistently with the weighted data. A convenient scaling factor is
\begin{equation}
\gamma
=
\frac{\left\|\mathbf{w}\odot\mathbf{f}\right\|_2}{N_s},
\qquad
\mathbf{f}=
\begin{bmatrix}
f(s_1) & f(s_2) & \cdots & f(s_{N_s})
\end{bmatrix}^{T},
\label{w}
\end{equation}
where \(\mathbf{w}=[w_1,\ldots,w_{N_s}]^T\), \(\odot\) denotes element-wise multiplication, and \(\gamma\) is used to scale both sides of Eq.~\eqref{rex2} in the augmented least-squares system.

After the relaxed scaling function has been estimated, its zeros are used as the relocated poles. Since
\begin{equation}
\sigma(s)
=
d_\sigma
\frac{\displaystyle\prod_{n=1}^{N_p}(s-z_n)}
     {\displaystyle\prod_{n=1}^{N_p}(s-a_n)},
\end{equation}
the zeros \(z_n\) of \(\sigma(s)\) are obtained from
\begin{equation}
\{z_n\}_{n=1}^{N_p}
=
\mathrm{eig}
\left(
\mathbf{A}
-
\mathbf{b}\,d_\sigma^{-1}\widetilde{\mathbf{C}}^{T}
\right),
\label{eq230}
\end{equation}
with \(\mathbf{A}\), \(\mathbf{b}\), and \(\widetilde{\mathbf{C}}^{T}\) as before.

The relocated poles are then assigned as \(a_n \leftarrow z_n\). The scaling function is only used for pole relocation; after convergence, it is discarded and the physical residues are recomputed from the original data with the final poles kept fixed.

\subsubsection{Fast Vector Fitting extension}

The fast VF extension reduces the computational cost of the pole-identification least-squares problem, especially for multi-output data, by avoiding the direct solution of one large dense system \cite{Deschrijver2008}. The key idea is to exploit the fact that the numerator residues are output-dependent, whereas the scaling-function residues are common to all fitted responses.

Let \(f_v(s_k)\) denote the measured frequency-domain samples of the \(v\)-th response, with \(v=1,\ldots,V\). For fixed current poles \(a_n\), define
\begin{equation}
\boldsymbol{\Phi}_{k,:}
=
\begin{bmatrix}
\dfrac{1}{s_k-a_1} &
\dfrac{1}{s_k-a_2} &
\cdots &
\dfrac{1}{s_k-a_{N_p}} &
1 &
s_k
\end{bmatrix},
\end{equation}
and
\begin{equation}
\boldsymbol{\Psi}_{k,:}
=
\begin{bmatrix}
\dfrac{1}{s_k-a_1} &
\dfrac{1}{s_k-a_2} &
\cdots &
\dfrac{1}{s_k-a_{N_p}} &
1
\end{bmatrix}.
\end{equation}
The output-dependent numerator coefficients are collected in
\begin{equation}
\boldsymbol{\theta}_v
=
\begin{bmatrix}
C_{v,1} & C_{v,2} & \cdots & C_{v,N_p} & d_v & e_v
\end{bmatrix}^{T},
\end{equation}
whereas the common scaling-function coefficients are collected in
\begin{equation}
\boldsymbol{\eta}
=
\begin{bmatrix}
\widetilde{C}_1 & \widetilde{C}_2 & \cdots & \widetilde{C}_{N_p} & d_\sigma
\end{bmatrix}^{T}.
\end{equation}
Furthermore, define
\begin{equation}
\mathbf{D}_v
=
\mathrm{diag}
\left(
f_v(s_1),f_v(s_2),\ldots,f_v(s_{N_s})
\right).
\end{equation}

The relaxed pole-identification problem for the \(v\)-th response can then be written compactly as
\begin{equation}
\boldsymbol{\Phi}\boldsymbol{\theta}_v
-
\mathbf{D}_v\boldsymbol{\Psi}\boldsymbol{\eta}
\approx
\mathbf{0}.
\label{eq:fast_partition}
\end{equation}
The unknown vector \(\boldsymbol{\theta}_v\) is local to each response, while \(\boldsymbol{\eta}\) is common to all responses and determines the pole relocation.

To eliminate the local numerator coefficients, a QR factorisation of \(\boldsymbol{\Phi}\) is performed:
\begin{equation}
\boldsymbol{\Phi}
=
\mathbf{Q}_v
\begin{bmatrix}
\mathbf{R}_v\\
\mathbf{0}
\end{bmatrix},
\qquad
\mathbf{Q}_v=
\begin{bmatrix}
\mathbf{Q}_{v,1} & \mathbf{Q}_{v,2}
\end{bmatrix}.
\label{eq:qr_fact}
\end{equation}
Premultiplying Eq.~\eqref{eq:fast_partition} by \(\mathbf{Q}_v^{H}\) (i.e., an Hermitian transpose) gives
\begin{equation}
\begin{cases}
\mathbf{R}_v\boldsymbol{\theta}_v
-
\mathbf{Q}_{v,1}^{H}\mathbf{D}_v\boldsymbol{\Psi}\boldsymbol{\eta}
\approx
\mathbf{0},
\\[4pt]
-
\mathbf{Q}_{v,2}^{H}\mathbf{D}_v\boldsymbol{\Psi}\boldsymbol{\eta}
\approx
\mathbf{0}.
\end{cases}
\end{equation}
The second equation contains only the common scaling-function coefficients \(\boldsymbol{\eta}\). Therefore, for all responses, the reduced systems can be stacked as
\begin{equation}
\begin{bmatrix}
\mathbf{Q}_{1,2}^{H}\mathbf{D}_1\boldsymbol{\Psi}\\
\mathbf{Q}_{2,2}^{H}\mathbf{D}_2\boldsymbol{\Psi}\\
\vdots\\
\mathbf{Q}_{V,2}^{H}\mathbf{D}_V\boldsymbol{\Psi}
\end{bmatrix}
\boldsymbol{\eta}
\approx
\mathbf{0}.
\end{equation}
The non-triviality constraint in Eq.~\eqref{rex2} is then appended to this reduced system, yielding a compact least-squares problem for \(\boldsymbol{\eta}\). Once \(\boldsymbol{\eta}\) has been estimated, the zeros of \(\sigma(s)\) are computed from Eq.~\eqref{eq230}, and the poles are relocated accordingly. After convergence, the final physical residues are computed with the relocated poles kept fixed, as described in the residue-identification stage of the basic VF algorithm.

Combining the relaxed pole-relocation formulation with this QR-based reduction gives the final formulation of the Fast and Relaxed Vector Fitting method. In the non-relaxed VF formulation, \(d_\sigma\) is fixed to one and the non-triviality constraint in Eq.~\eqref{rex2} is not required. In the non-fast formulation, the full least-squares system is solved directly, without the QR-based elimination of the local numerator coefficients.

\subsection{Natural Excitation Technique}

NExT estimates auto- and cross-correlation functions from output-only vibration data. Under suitable assumptions, these correlation functions have the same free-decay structure as impulse response functions (IRFs), up to an unknown scaling factor. They can therefore be used within system identification (SI) methods, such as ERA, to estimate modal parameters \cite{Dessena2024f}. The technique was first introduced in output-only settings to identify wind turbines during operational conditions \cite{GeorgeIII1993} and has since been adopted in a wide range of applications where excitation cannot be measured.

In fact, in SI, structural responses are typically acquired together with the applied input forces, which is not feasible in output-only testing. When NExT is used, the unmeasured excitation is instead modelled as a random, stationary, and broadband process, capable of activating the relevant modes of a linear time-invariant (LTI) system \cite{Peeters1999}. 

In discrete time, a stochastic LTI description with zero-mean noise can be written as
\begin{equation}
    \mathbf{x}_{k+1} = \mathbf{A} \mathbf{x}_k + \mathbf{w}_k, 
    \qquad 
    \mathbf{y}_k = \mathbf{C}_{LTI} \mathbf{x}_k + \mathbf{v}_k ,
\end{equation}
where \(k \in \mathbb{N}\) is the discrete-time index, \(\mathbf{x}_k \in \mathbb{R}^{n}\) is the state vector, \(\mathbf{y}_k \in \mathbb{R}^{p}\) is the measured output vector, \(\mathbf{A} \in \mathbb{R}^{n \times n}\) is the state-transition matrix, \(\mathbf{C}_{LTI} \in \mathbb{R}^{p \times n}\) is the output matrix, and \(\mathbf{w}_k\) and \(\mathbf{v}_k\) are the process and measurement noise terms, respectively.

Denoting the expectation operator by \(E[\cdot]\), it is assumed that the state is zero-mean, \(E[\mathbf{x}_k]=0\), with covariance \(E[\mathbf{x}_k \mathbf{x}_k^{T}]=\Sigma\), where \(\Sigma \in \mathbb{R}^{n \times n}\) is the state covariance matrix.
Marked deviations from stationarity (i.e., strongly time-varying operating conditions) compromise the correlation estimates and can reduce the accuracy of the resulting IRFs.

For positive time lags, the output correlation matrix is
\begin{equation}
\mathbf{R}_{yy}(\ell)
=
E\left[
\mathbf{y}_{k+\ell}\mathbf{y}_{k}^{T}
\right]
=
\mathbf{C}_{LTI}\mathbf{A}^{\ell}\mathbf{G},
\qquad \ell \geq 0,
\end{equation}
where
\begin{equation}
\mathbf{G}=E\left[\mathbf{x}_k\mathbf{y}_k^{T}\right].
\end{equation}
This expression contains the same state-transition matrix \(\mathbf{A}\) as the impulse response of the system. Therefore, the output correlation functions contain the same modal poles as the underlying structure and can be interpreted as impulse-response-like free decays.

For two measured output signals \(y_i(t)\) and \(y_j(t)\), the continuous-time cross-correlation function is estimated as
\begin{equation}
    R_{y_i y_j}(\tau) =
    \lim_{T_s \to \infty}
    \frac{1}{T_s}
    \int_{0}^{T_s}
    y_i(t)\,y_j(t+\tau)\,dt ,
\end{equation}
where \(R_{y_i y_j}(\tau)\) is the cross-correlation function, \(\tau\) is the time lag, and \(T_s\) is the observation time. In practice, the correlation is estimated over a finite time window, so insufficient data length can also reduce the quality of the estimated free decays.

In this study, the NExT-derived correlation functions are used directly by ERA as free-decay response data, while their frequency-domain representation is obtained via FFT and supplied to FRVF. The resulting frequency-domain functions are therefore NExT-derived spectral functions, rather than directly measured input-output FRFs.

Nevertheless, as said, under the assumption of broadband and stationary excitation, \(R_{y_i y_j}(\tau)\) provides an equivalent impulse-response-type signature (up to a scaling factor), allowing modal identification without direct input measurements \cite{GeorgeIII1993}.

\subsection{NExT-FRVF\label{sec:nextfrvf}}
To complete the Methodology description, as its final step, NExT is combined with FRVF by first estimating the IRF directly from time-domain output measurements and then mapping this IRF to the frequency domain via Fast Fourier Transform (FFT). The resulting FRF provides the frequency-domain data required by FRVF to identify the system modes. The complete workflow adopted in this study, from correlation-based IRF estimation to iterative pole update and modal extraction, is reported in \cref{alg:next-frvf}.

\begin{algorithm}[htp!]
\caption{Implementation of the NExT--FRVF method}
\label{alg:next-frvf}
\begin{algorithmic}[1]
\State \textbf{Input:} response signals \(y(t)\), model order \(N_p\), number of pole-relocation iterations \(N_{\mathrm{iter}}\), reference channels, maximum correlation lag \(\tau_{\max}\), and sampling frequency \(f_s\).
\State \textbf{Output:} identified poles \(p_n\), physical residues \(R_n\), coefficients \(d\) and \(e\), and modal parameters (natural frequencies \(f_n\), damping ratios \(\zeta_n\), and mode shapes \(\phi_n\)).

\State Estimate the auto- and cross-correlation functions \(R_{ij}(\tau)\) from the measured output channels.
\State Interpret the correlation functions as IRF-like free decays (according to the common NExT assumptions).
\State Compute the NExT-derived experimental data \(f(s_k)\) by FFT, in an FRF-like fashion (again, following the same standard NExT assumptions).
\State Initialise the trial poles \(a_n^{(0)}\) using Eq.~\eqref{init}.

\For{\(i=1,\ldots,N_{\mathrm{iter}}\)}
    \State Formulate the pole-identification LS problem using the current poles \(a_n^{(i-1)}\), Eq.~\eqref{eq220}.
    \State Solve the LS problem using the fast QR-based formulation, Eq.~\eqref{eq:qr_fact}.
    \State Estimate the scaling-function residues \(\widetilde{C}_n\) and, in the relaxed formulation, \(d_\sigma\).
    \State Compute the zeros \(z_n\) of \(\sigma(s)\) using Eq.~\eqref{eq230}.
    \State Relocate the poles: \(a_n^{(i)} \leftarrow z_n\).
\EndFor

\State Set the final identified poles as \(p_n \leftarrow a_n^{(N_{\mathrm{iter}})}\).
\State With \(p_n\) fixed, solve the residue-identification LS problem to estimate \(R_n\), \(d\), and \(e\).
\State Compute the modal parameters:
\[
\omega_n=|p_n|,
\qquad
f_n=\frac{\omega_n}{2\pi},
\qquad
\zeta_n=-\frac{\mathrm{Re}(p_n)}{|p_n|}.
\]
\State Estimate and normalise the mode shapes \(\phi_n\) from the residue vectors.
\end{algorithmic}
\end{algorithm}

\section{Numerical case study\label{sec:num}}

\begin{figure}[htp!]
\centering
\resizebox{\textwidth}{!}{  
\makebox[\columnwidth][c]{%
  \begin{minipage}[c]{0.80\columnwidth}
    \centering
    \begin{subfigure}[c]{0.725\linewidth} 
      \centering
      \includegraphics[width=\linewidth]{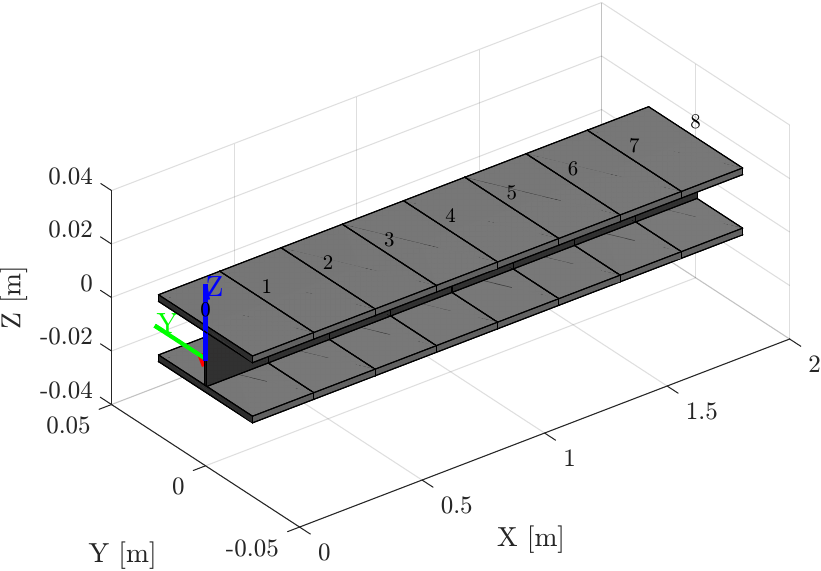}
      \caption{}
      \label{fig:beam_main}
    \end{subfigure}%
    \hspace{0.025\linewidth}%
    \begin{subfigure}[c]{0.25\linewidth}
      \centering
      \includegraphics[width=\linewidth]{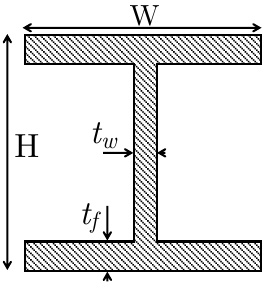}
      \caption{}
      \label{fig:beam_dims}
    \end{subfigure}
    \caption{Numerical model and I-beam cross-section geometry used in the validation study: (a) Euler-Bernoulli 2D beam model; (b) relevant dimensions of the I-beam.}
    \label{fig:main_beam_figure}
  \end{minipage}%
  \hspace{0.02\columnwidth}%
  \begin{minipage}[c]{0.18\columnwidth}
    \centering
    \captionof{table}{I-beam main dimensions values.}
    \label{tab:beam_dims_values}
    \vspace{2mm} 
    \renewcommand{\arraystretch}{1.2}
    \setlength{\tabcolsep}{4pt}
    \begin{tabular}{lc}
      \toprule
      \textbf{Dimensions} & \textbf{Value [m]} \\
      \midrule
      H     & 0.025  \\
      W     & 0.050   \\
      $t_w$ & 0.0015 \\
      $t_f$ & 0.0025 \\
      \bottomrule
    \end{tabular}
  \end{minipage}%
}}
\end{figure}

The numerical model to preliminarily validate the proposed NExT-FRVF method is an Euler--Bernoulli (EB) cantilever beam (\cref{fig:beam_main}) with an I-shaped cross-section. The beam, clamped at the root (corresponding to node 0), spans 2 m and its sectional dimensions are displayed within \cref{fig:beam_dims,tab:beam_dims_values}. It is made of standard Aluminium, yielding the following properties: (Density) $\rho =$ 2700 kgm \textsuperscript{-3} and (Young's modulus) $E =$ 70 GPa.

The beam is discretised using eight EB elements ($z$ transverse displacements). The damping matrix is built from the uncoupled modal damping assumption, imposing a 3\% damping ratio across all modes, while stiffness and mass matrices follow the classic EB formulation. The element mass ($\mathbf{M}$) and stiffness matrices ($\mathbf{K}$) used are shown in \cref{eq:mass,eq:stiff}, in \ref{sec:appA}.

A vertical impulse force of 1 N is applied at node 1 at \(t=0\,\mathrm{s}\), lasting \(f_s^{-1}\) seconds. The simulation sampling frequency is \(f_s = 3600\, \mathrm{Hz}\) and lasts \(T_s = 30\, \mathrm{s}\), capturing frequencies up to 1800 Hz, thus including all modes of interest (the first eight) according to the Nyquist criterion.

\begin{figure}[htp!]
  \centering
        \begin{subfigure}[c]{.49\textwidth}
          \centering
          \includegraphics[width=.9\textwidth]{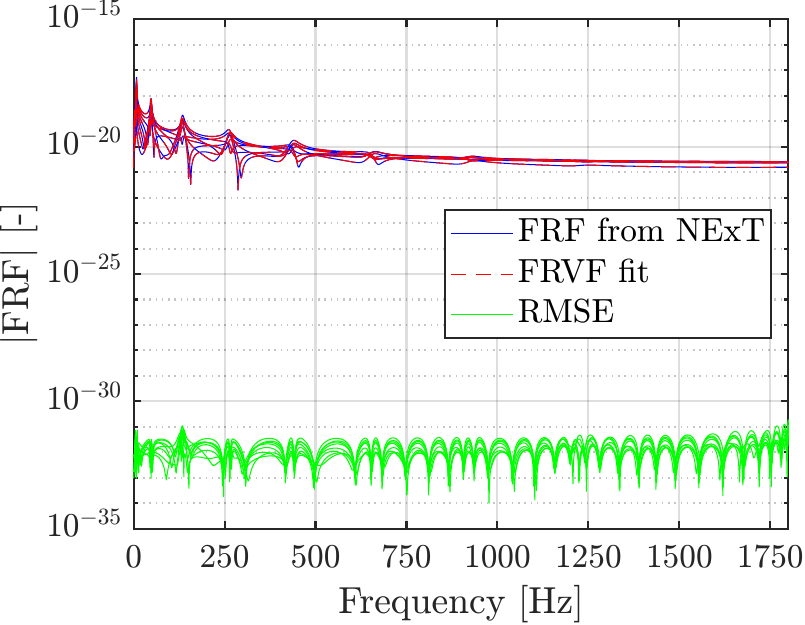}
          \caption{}
          \label{fig:sub1}
        \end{subfigure}
        \begin{subfigure}[c]{.49\textwidth}
          \centering
          \includegraphics[width=.9\textwidth]{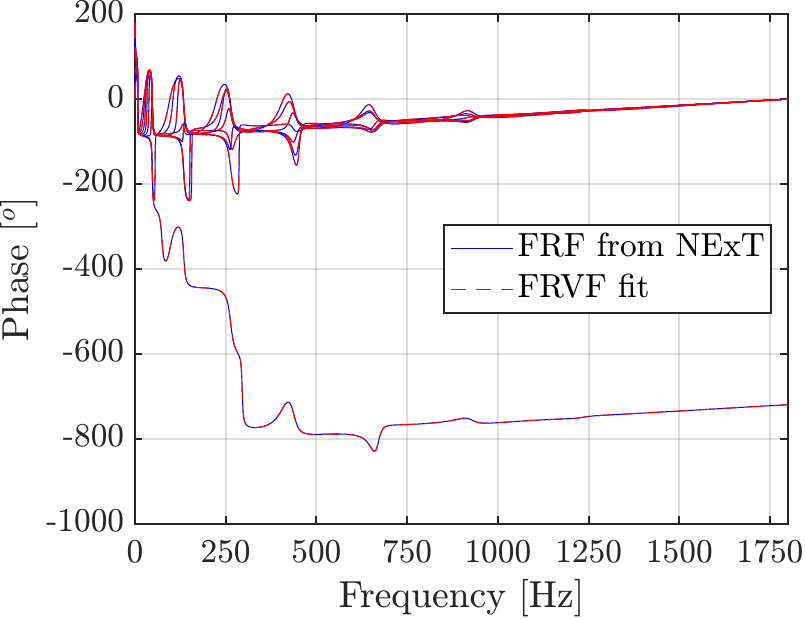}
          \caption{}
        \end{subfigure}
  \caption{FRVF approximation in noiseless case.}
  \label{f34}
\end{figure}

Using the first channel displacement as reference, the IRF is obtained via NExT, as shown in \cref{alg:next-frvf}.\Cref{f34} shows the actual NExT-derived FRF data (solid blue), the model approximation ${(\sigma H)_{\text{fit}}(s)}/{\sigma_{\text{fit}}(s)}$ (dashed red) using a minimum order realisation (corresponding to an FRVF model of order equal to two times the number of expected modes) and the Root Mean Square Error (RMSE -- green), for the FRF magnitude only case. It is worth noting that, using 5 iterations, the FRVF method yields an almost perfect approximation of the original FRF since the global RMSE is orders of magnitude less than the FRF amplitude.

After this preliminary evaluation, the modal parameters are extracted from the numerical dataset. To this end, higher-order realisations of FRVF, and the benchmark method ERA, are considered, with stabilisation diagrams used to identify physically meaningful and stable modes. Stabilisation diagrams display recurrent poles that represent physical modes across model orders \cite{Scionti2005}. In this case, the diagrams are constructed by varying the model order $k$ from $k_{\min}=8$ to $k_{\max}=100$ with increment $\Delta k=2$, yielding the extraction set $k=\{k_{\min}:\Delta k:k_{\max}\}$. Hard screening is imposed through the admissible damping range $\zeta_n\in[\zeta_{\min},\zeta_{\max}]=[0.01,0.05]$ and the natural-frequency range $f_n\in[f_{\min},f_{\max}]=[0,1300]\,\mathrm{Hz}$. Stable poles are then selected using soft criteria based on frequency stabilisation $\Delta f_{\mathrm{stab}}=0.01$, damping stabilisation $\Delta\zeta_{\mathrm{stab}}=0.3$, and consistency of the mode shape enforced by $\mathrm{MAC}_{\mathrm{stab}}=0.95$. Additional screening is introduced through the tolerance parameter $\epsilon=0.1$ and by requiring consistency over $N_{\mathrm{MAC}}=5$ consecutive model orders.
Although the diagrams are not shown here for brevity, at higher model orders, NExT-FRVF tends to produce more spurious high-frequency modes than NExT-ERA, as the system attempts to identify modes in frequency ranges that are only weakly excited; therefore, the maximum model order $n_{max}$ should be selected with care. 

\Cref{tab:results_gray_lines} reports the modal parameters identified by NExT-FRVF and NExT-ERA, together with the analytical results obtained by eigenanalysis and the corresponding errors. Mode shapes $\mathbf{\phi}_n$ are presented in terms of their agreement, via the modal assurance criterion (MAC), with the analytical solution.

\begin{table}[htp!]
\centering
\small
\renewcommand{\arraystretch}{1.2}
\setlength{\tabcolsep}{5pt}
\caption{Results obtained by analytical, NExT-ERA, and NExT-FRVF methods from the numerical case. Below each value of NExT-ERA and NExT-FRVF $f_n$ ($\omega_n=2\pi f_n$), $\zeta_n$, the corresponding percentage differences with respect to the analytical results are reported.}
\begin{tabularx}{\textwidth}{l *{3}{>{\centering\arraybackslash}X} *{3}{>{\centering\arraybackslash}X} *{2}{>{\centering\arraybackslash}X}}
\toprule
 & \multicolumn{3}{c}{\textbf{Natural Frequency} [Hz]}
 & \multicolumn{3}{c}{\textbf{Damping Ratio} [--]}
 & \multicolumn{2}{c}{\textbf{MAC} (diagonal values) [--]} \\
\midrule
Mode & Analytical & NExT-ERA & NExT-FRVF
     & Analytical & NExT-ERA & NExT-FRVF
     & NExT-ERA & NExT-FRVF \\
\midrule
1 & 7.648 & 7.651 & 7.651 & 0.030 & 0.030 & 0.030 & 1.000 & 1.000 \\
  &       & (0.04\%) & (0.04\%) &       & (1.34\%) & (1.35\%) &  &  \\
2 & 47.930 & 47.952 & 47.952 & 0.030 & 0.030 & 0.030 & 1.000 & 1.000 \\
  &       & (0.04\%) & (0.04\%) &       & (0.21\%) & (0.36\%) &  &  \\
3 & 134.277 & 134.337 & 134.337 & 0.030 & 0.030 & 0.030 & 1.000 & 0.998 \\
  &       & (0.04\%) & (0.04\%) &       & (0.46\%) & (0.13\%) &  &  \\
4 & 263.558 & 263.676 & 263.677 & 0.030 & 0.030 & 0.030 & 1.000 & 0.998 \\
  &       & (0.04\%) & (0.04\%) &       & (0.21\%) & (0.07\%) &  &  \\
5 & 437.222 & 437.396 & 437.419 & 0.030 & 0.030 & 0.030 & 1.000 & 0.991 \\
  &       & (0.04\%) & (0.04\%) &       & (0.18\%) & (0.04\%) &  &  \\
6 & 657.098 & 657.370 & 657.392 & 0.030 & 0.030 & 0.030 & 0.998 & 0.993 \\
  &       & (0.04\%) & (0.04\%) &       & (0.16\%) & (0.02\%) &  &  \\
7 & 925.002 & 926.896 & 925.419 & 0.030 & 0.031 & 0.030 & 0.953 & 0.978 \\
  &       & (0.20\%) & (0.05\%) &       & (3.05\%) & (0.02\%) &  &  \\
8 & 1228.851 & 1229.407 & 1229.405 & 0.030 & 0.030 & 0.030 & 0.715 & 0.933 \\
  &       & (0.05\%) & (0.05\%) &       & (0.02\%) & (0.01\%) &  &  \\
\bottomrule
\end{tabularx}
\label{tab:results_gray_lines}
\end{table}

The results in \cref{tab:results_gray_lines} show strong agreement between NExT-FRVF and the analytical solution across all eight modes, with errors in natural frequency ($f_n=\omega_n/2\pi$) not exceeding 0.05\% and damping ratio errors that remain below 1.35\% for all identified modes. NExT-ERA produces comparable $f_n$ and $\omega_n$ estimates for Modes~1--6, but exhibits a marked degradation for Mode~7: The identified $f_n$ deviates by 0.20\%, the $\zeta_n$ error reaches 3.05\%, and the MAC value falls to 0.953, against the NExT-FRVF values of 0.05\%, 0.02\%, and 0.978, respectively. For Mode~8, the NExT-FRVF-identified $\mathbf{\phi}_8$ achieves a MAC of 0.933, whilst the NExT-ERA counterpart yields only 0.715. Taken together, these results show that, despite the tendency of NExT-FRVF to produce spurious high-frequency poles at elevated model orders, it yields superior frequency, damping, and mode shape accuracy for the higher modes relative to NExT-ERA in the present numerical case.

The FRVF parameters selected for this study are as follows: No frequency weighting scheme and 5 iterations. These serve as the basis for the parametric analysis of noise effects, weighting schemes, and iteration count.

\subsection{Analysis of noise effect}

\begin{figure}[htp!]
\centering
% --- Row 1: NExT-ERA (Top) ---
\begin{subfigure}[t]{0.32\textwidth}
\centering
\includegraphics[width=\textwidth]{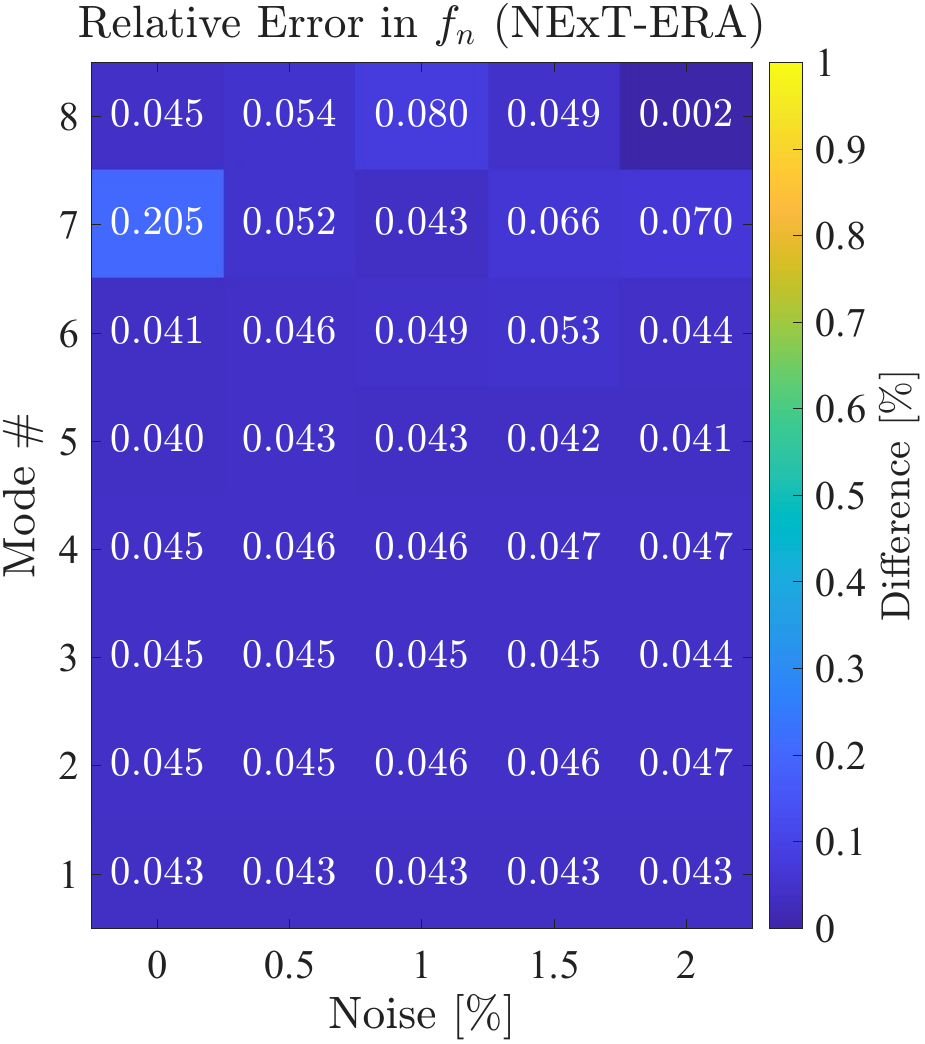}
\caption{}
\label{fig:color2a}
\end{subfigure}
\hfill
\begin{subfigure}[t]{0.32\textwidth}
\centering
\includegraphics[width=0.985\textwidth]{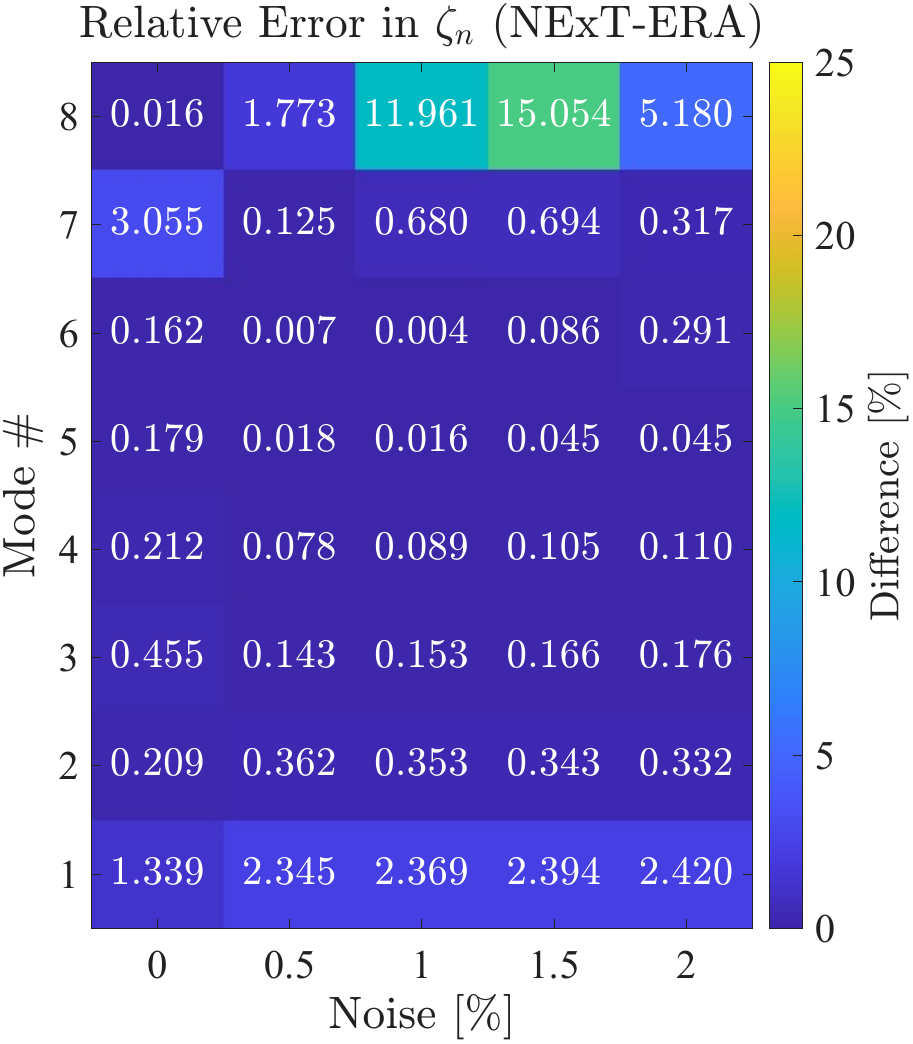}
\caption{}
\label{fig:color2b}
\end{subfigure}
\hfill
\begin{subfigure}[t]{0.32\textwidth}
\centering
\includegraphics[width=0.985\textwidth]{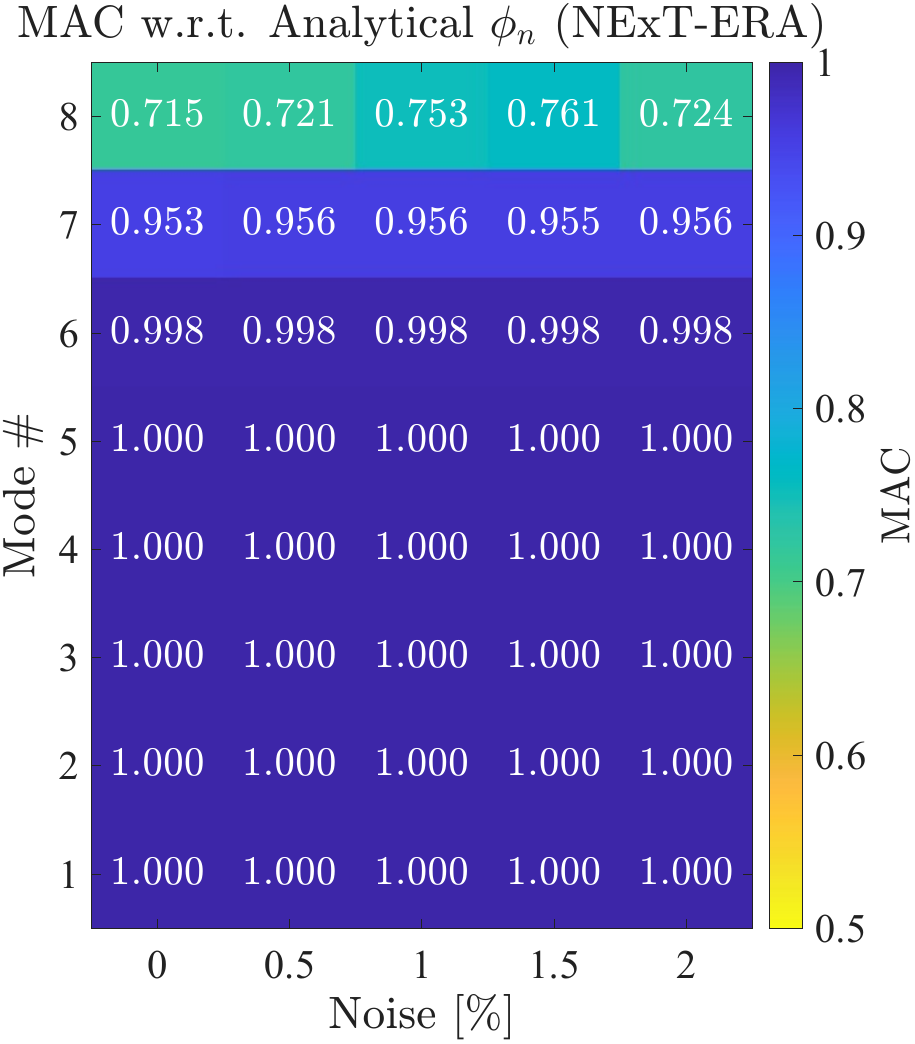}
\caption{}
\label{fig:color2c}
\end{subfigure}

\vspace{0.4cm} % Vertical separation between rows

% --- Row 2: NExT-FRVF (Bottom) ---
\begin{subfigure}[t]{0.32\textwidth}
\centering
\includegraphics[width=\textwidth]{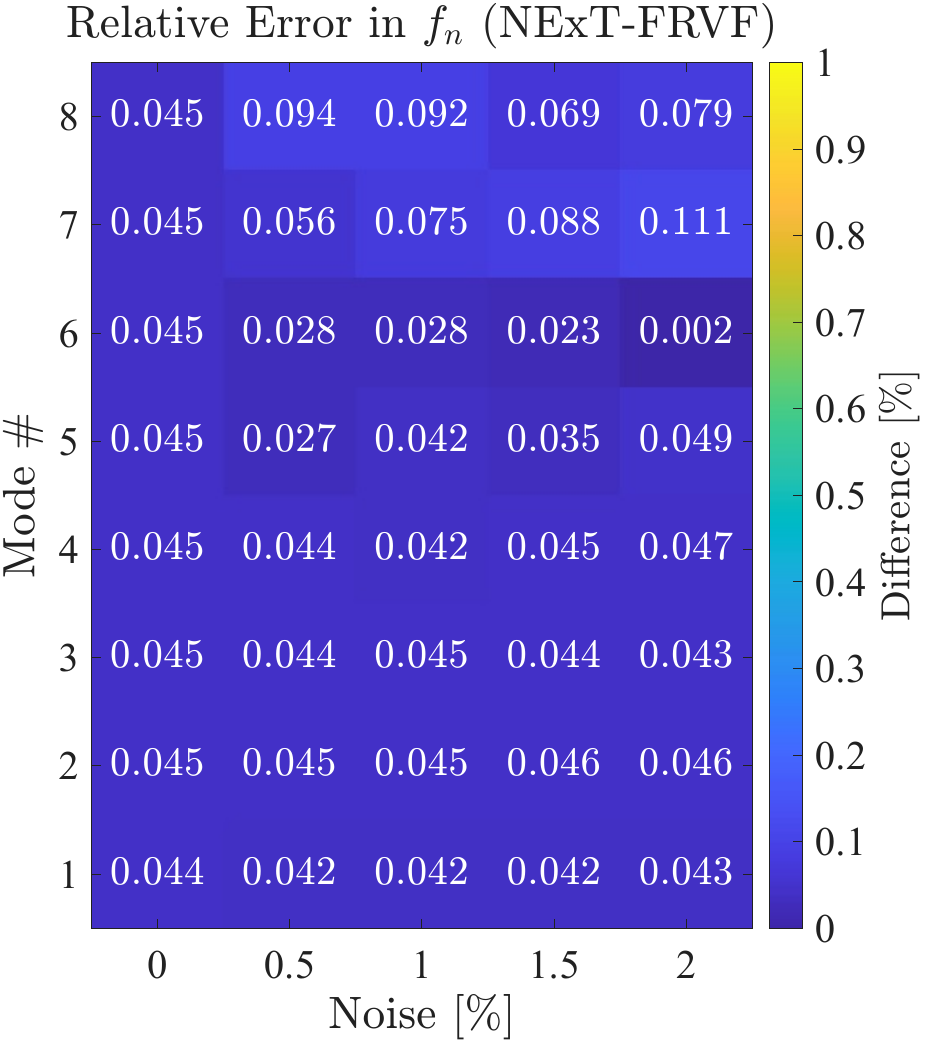}
\caption{}
\label{fig:color2d}
\end{subfigure}
\hfill
\begin{subfigure}[t]{0.32\textwidth}
\centering
\includegraphics[width=.985\textwidth]{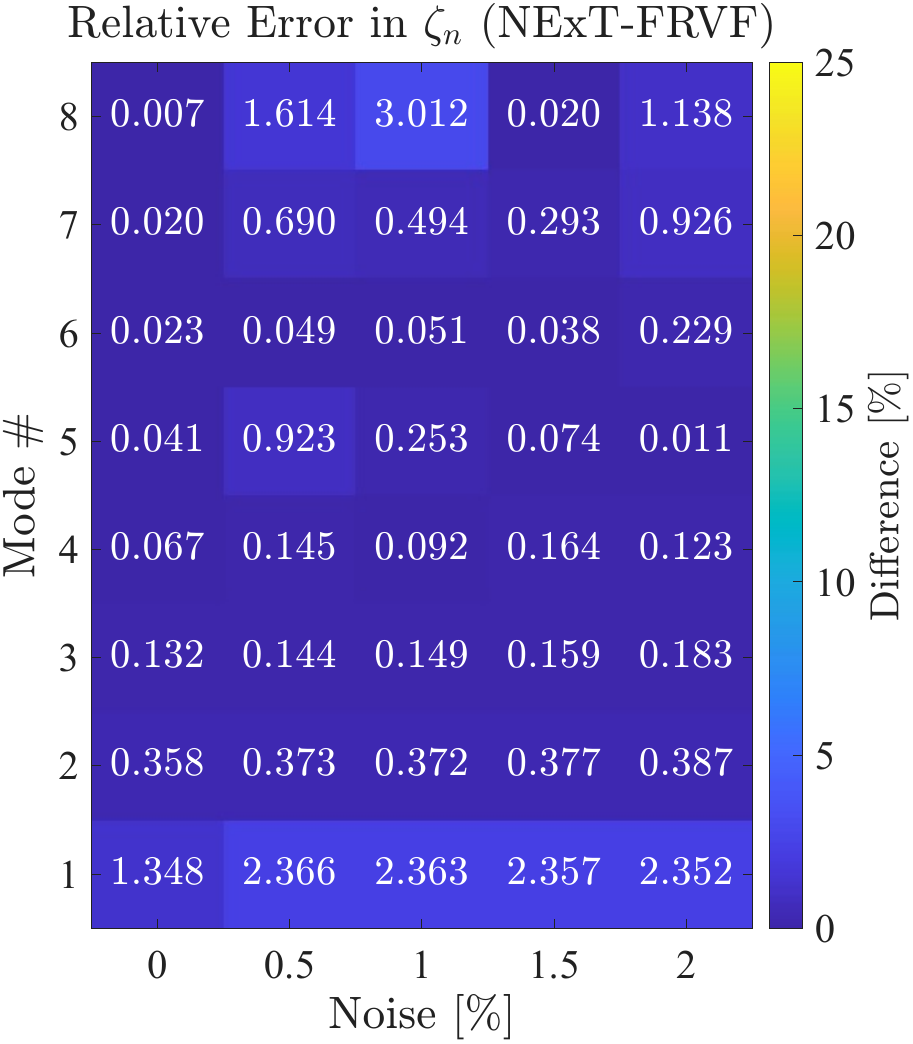}
\caption{}
\label{fig:color2e}
\end{subfigure}
\hfill
\begin{subfigure}[t]{0.32\textwidth}
\centering
\includegraphics[width=0.985\textwidth]{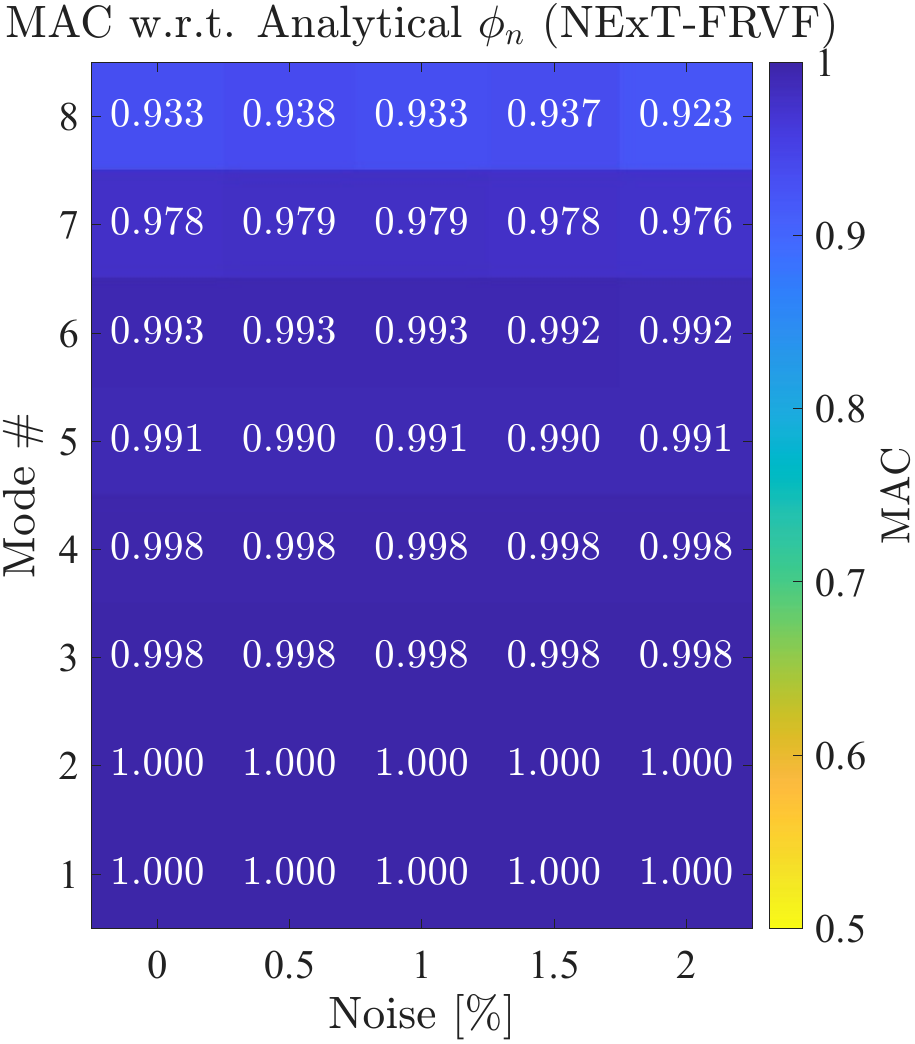}
\caption{}
\label{fig:color2f}
\end{subfigure}

\caption{Noise effect on identified modal parameters relative to the analytical solution: (a, d) natural frequency error, (b, e) damping ratio error, and (c, f) MAC values for NExT-ERA and NExT-FRVF, respectively.}
\label{fig:color2}
\end{figure}

The effect of measurement noise on the proposed method is assessed by introducing Additive White Gaussian Noise (AWGN), obtained by scaling a random array by the standard deviation of each output signal element and a prescribed noise percentage. The noise levels considered are $0$ (i.e., noise-free conditions), $0.5$, $1$, $1.5$, and $2\%$. Thus, the processing and identification described for the noise-free scenario are repeated for the four noise cases. This assessment is relevant in that noise can hinder convergence to an accurate fit, primarily by promoting the identification of spurious poles that capture noise components rather than the underlying structural dynamics. The evolution of $f_n$, $\omega_n$, and $\mathbf{\phi}_n$ identification as a function of the noise level is summarised in \Cref{fig:color2}.

Overall, the two methods show comparable behaviour across noise levels for \(f_n\) identification, with percentage differences remaining below \(0.12\%\) for all modes successfully identified by NExT-FRVF and below \(0.21\%\) for those identified by NExT-ERA. 
The estimates of the damping ratio \((\zeta_n)\) are more sensitive to noise, as expected, with the largest errors occurring mainly for the higher modes.

For NExT-ERA, Mode~8 is the most affected in terms of damping-ratio error, reaching \(11.961\%\), \(15.054\%\), and \(5.180\%\) at noise levels of \(1\%\), \(1.5\%\), and \(2\%\), respectively. For comparison, Mode~1 shows a moderate increase in damping error, reaching \(2.394\%\) and \(2.420\%\) at \(1.5\%\) and \(2\%\) noise, respectively. 
Notably, some modes (e.g., Mode~7) do not exhibit a monotonic deterioration. Conversely, NExT-FRVF exhibits consistent damping estimates for Mode~1 in the noisy cases, with errors remaining approximately \(2.35\%\). It also remains accurate for several higher modes: For example, Mode~6 shows damping errors of \(0.038\%\) and \(0.229\%\) at \(1.5\%\) and \(2\%\) noise, respectively.

Also for Mode~8, NExT-FRVF provides consistently lower damping-ratio errors than NExT-ERA across all noise levels. The improvement is particularly evident at \(1\%\), \(1.5\%\), and \(2\%\) noise, where the \(\zeta_n\) error decreases from \(11.961\%\), \(15.054\%\), and \(5.180\%\) for NExT-ERA to \(3.012\%\), \(0.020\%\), and \(1.138\%\) for NExT-FRVF, respectively.  Thus, the differences between NExT-ERA and NExT-FRVF become more pronounced for the higher modes and under noisy conditions.

The differences between NExT-ERA and NExT-FRVF are also evident in the mode shape estimates: Again for Mode~8, the NExT-ERA MAC values range from \(0.715\) to \(0.761\), indicating reduced mode shape consistency, while the NExT-FRVF MAC values remain between \(0.923\) and \(0.938\), much higher. These results indicate that NExT-FRVF is more robust to noise than NExT-ERA for the higher modes, especially in terms of damping-ratio and mode-shape estimates.

\subsection{Analysis of weight effect}
Weighting adjusts the FRVF approximation by tuning the relative importance of frequency regions in the weighted LS problem. Three standard schemes are considered \cite{Gustavsen2008}:

\begin{itemize}
    \item \textit{No weighting}, used by default;
    \item \textit{Weak inverse weighting}, which provides a moderate emphasis of low-magnitude regions;
    \item \textit{Strong inverse weighting}, which further increases the emphasis of low-magnitude regions but may over-amplify noise contributions.
\end{itemize}
\Cref{tab36} summarises the three possible variants for the weighting function.

\begin{table}[htp!]
\centering
\small
\renewcommand{\arraystretch}{1.3}
\setlength{\tabcolsep}{4pt}
\begin{tabular}{lcc}
\toprule
\textbf{No weight} &  \textbf{Weak inverse weight} & \textbf{Strong inverse weight} \\
\midrule
$weight = 1$  & $weight = \frac{1}{\sqrt{\left | signal \right |}}$ & $weight = \frac{1}{\left | signal \right |}$ \\
\bottomrule
\end{tabular}
\caption{FRVF weighting schemes.}
\label{tab36}
\end{table}

This analysis adopts the 2\% noise case as a representative scenario, which makes it suitable for assessing the effect of weighting on both the quality of fitting and the robustness of the estimated modal parameters. The NExT-FRVF identification is carried out at the lowest order ($N_p=16$), and the RMSE between the original and fitted FRF is used as the evaluation index. The resulting RMSE values are reported in \cref{tabw1}.

\begin{table}[htp!]
\centering
\small
\renewcommand{\arraystretch}{1.3}
\setlength{\tabcolsep}{5.5pt}
\begin{tabular}{lccc}
\toprule
\textbf{Weight} & \textbf{No weight} & \textbf{Weak inverse} & \textbf{Strong inverse} \\
\midrule
\textbf{RMSE} & $7.194\cdot10^{-23}$ & $9.497\cdot10^{-23}$ & $6.503\cdot10^{-22}$ \\
\bottomrule
\end{tabular}
\caption{RMSE comparison for different weighting schemes for the final iteration of the $noise=2\%$ numerical case.}
\label{tabw1}
\end{table}

The RMSE increases from $7.194 \times 10^{-23}$ under unit weighting to $6.503 \times 10^{-22}$ under strong inverse weighting, a difference of approximately one order of magnitude; nevertheless, all three values remain many orders of magnitude below the signal level, confirming that the rational approximation is effectively exact under all schemes in absolute terms. However, the global RMSE is dominated by the high-magnitude frequency content associated with the well-excited lower modes and is therefore not a sole reliable indicator of the identification quality for weakly excited modes. Unit weighting concentrates the fitting effort on these high-magnitude regions, minimising the absolute RMSE but proportionally minimising the representational capacity to the low-magnitude spectral content where Modes~7 and~8 reside. Inverse weighting redistributes this effort by amplifying the relative contribution of low-magnitude regions to the LS problem, necessarily increasing the absolute RMSE whilst improving pole resolution in those regions. The two objectives -- minimising global absolute fitting error and accurately characterising weakly excited modes -- are therefore not equivalent, and the RMSE alone is an insufficient criterion for weighting scheme selection when weak modes are of primary interest.

The modal parameter estimates, reported in \cref{tab:results_weighted}, reveal a clear dependence on the weighting scheme for the higher, weakly excited modes, whereas Modes~1--6 remain unaffected. For these modes, $f_n$ deviations are below 0.06\%, $\zeta_n$ errors below 2.40\%, and MAC values $\geq 0.990$  in all three schemes, with no consistent directional trend across weighting formulations, confirming that spectral rebalancing does not have a material effect on well-excited modes. For Mode~7, the inverse weighting substantially reduces the $\zeta_n$ error: From 10.18\% under unit weighting to 2.19\% and 1.63\% under weak and strong inverse weighting, respectively, whilst $f_n$ and MAC estimates remain comparable across schemes. Mode~8 is not recovered under unit weighting; weak inverse weighting identifies a candidate at 1239.190 Hz, but with MAC$= 0.118$, indicating an incorrect $\mathbf{\phi}_n$, whilst strong inverse weighting yields a physically meaningful identification at 1229.385 Hz, with MAC$= 0.870$.

\begin{table}[htp!]
\centering
\small
\caption{Results obtained by NExT-FRVF with no weight, weak inverse weight, and strong inverse weight in the numerical case study with $noise=2\%$. Below each value of $f_n$ and $\zeta_n$, the corresponding percentage differences with respect to the analytical results are reported.}
\renewcommand{\arraystretch}{1.2}
\setlength{\tabcolsep}{5pt}
\begin{tabularx}{\textwidth}{l *{3}{>{\centering\arraybackslash}X} *{3}{>{\centering\arraybackslash}X} *{3}{>{\centering\arraybackslash}X}}
\toprule
 & \multicolumn{3}{c}{\textbf{Natural Frequency} [Hz]}
 & \multicolumn{3}{c}{\textbf{Damping Ratio} [--]}
 & \multicolumn{3}{c}{\textbf{MAC} (diagonal values) [--]} \\
\midrule
Mode & No weight & Weak inverse & Strong inverse
     & No weight & Weak inverse & Strong inverse
     & No weight & Weak inverse & Strong inverse \\
\midrule

1 & 7.651 & 7.651 & 7.651 & 0.029 & 0.029 & 0.029 & 1.000 & 1.000 & 1.000 \\
  & (0.04\%) & (0.04\%) & (0.04\%) & (2.40\%) & (2.31\%) & (1.98\%) &  &  &  \\

2 & 47.952 & 47.952 & 47.953 & 0.030 & 0.030 & 0.030 & 1.000 & 1.000 & 1.000 \\
  & (0.05\%) & (0.05\%) & (0.05\%) & (0.37\%) & (0.26\%) & (0.17\%) &  &  &  \\

3 & 134.334 & 134.337 & 134.345 & 0.030 & 0.030 & 0.030 & 0.998 & 0.998 & 0.998 \\
  & (0.04\%) & (0.04\%) & (0.05\%) & (0.26\%) & (0.09\%) & (0.27\%) &  &  &  \\

4 & 263.681 & 263.682 & 263.709 & 0.030 & 0.030 & 0.030 & 0.998 & 0.998 & 0.998 \\
  & (0.05\%) & (0.05\%) & (0.06\%) & (0.04\%) & (0.11\%) & (0.17\%) &  &  &  \\

5 & 437.344 & 437.432 & 437.487 & 0.030 & 0.030 & 0.030 & 0.991 & 0.990 & 0.990 \\
  & (0.03\%) & (0.05\%) & (0.06\%) & (0.24\%) & (0.06\%) & (0.03\%) &  &  &  \\

6 & 657.523 & 657.251 & 657.323 & 0.030 & 0.030 & 0.030 & 0.992 & 0.993 & 0.993 \\
  & (0.06\%) & (0.02\%) & (0.03\%) & (0.18\%) & (0.32\%) & (0.69\%) &  &  &  \\

7 & 927.766 & 926.106 & 926.151 & 0.033 & 0.031 & 0.030 & 0.984 & 0.977 & 0.979 \\
  & (0.30\%) & (0.12\%) & (0.12\%) & (10.18\%) & (2.19\%) & (1.63\%) &  &  &  \\

8 & Not found & 1239.190 & 1229.385 & Not found & 0.043 & 0.039 & Not found & 0.118 & 0.870 \\
  &  & (0.84\%) & (0.04\%) &  & (43.54\%) & (28.72\%) &  &  &  \\
\bottomrule
\end{tabularx}
\label{tab:results_weighted}
\end{table}

These results are consistent with the mechanism described above: Inverse weighting amplifies the influence of low-magnitude spectral regions, enabling the pole relocation process to resolve weakly excited modes despite the elevated noise level, at the cost of increased absolute RMSE that remains negligible relative to the signal.

\subsection{Analysis of the effect of the number of iterations}
\begin{figure}[htp!]
\centering
{\includegraphics[width=0.5\textwidth]{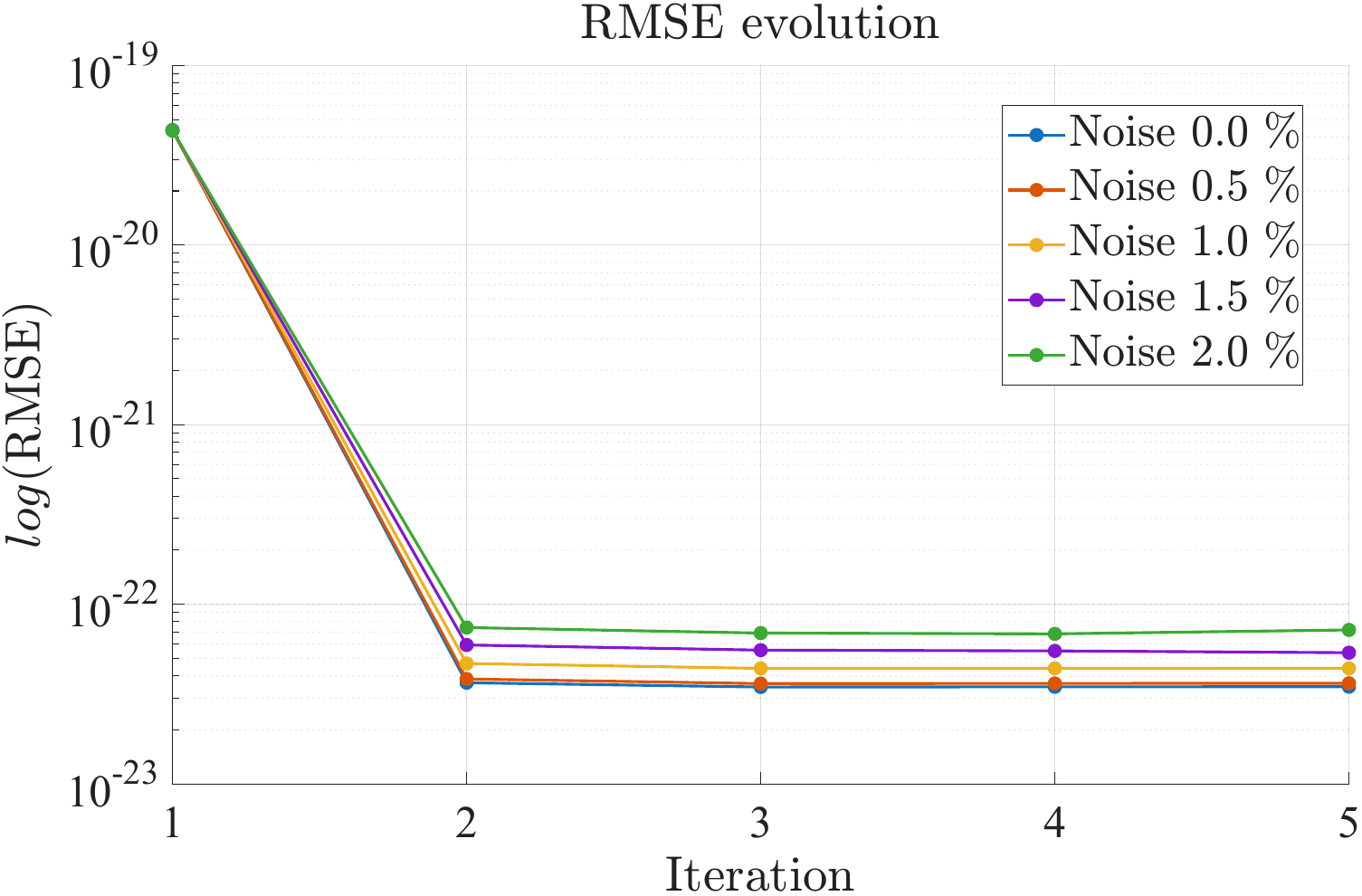}}
\caption{RMSE evolution across iterations for each noise scenario
         considered in the numerical case study.}
\label{rms}
\end{figure}

The number of iterations in the FRVF approach affects the convergence of the estimated poles and, consequently, the reliability of the extracted modal parameters. At each iteration, the pole set is reallocated to progressively refine the rational approximation, at the cost of increased computational effort. Keeping the iteration count as low as possible is therefore desirable, provided that fitting accuracy and modal estimates remain satisfactory. In this analysis, the 2\% noise case is considered at the lowest model order ($N_p=16$). The evolution of the RMSE fitting error across iterations and noise levels is reported in \cref{rms}. 

In noise-free conditions, a single round ($N_\mathrm{iter} = 2$) of pole relocation is practically sufficient, already returning a very low RMSE. As the noise level increases, the RMSE curves begin to diverge across iterations, following an approximately linear trend with the noise level, with the fifth iteration yielding a measurably lower error than the preceding one at 2\% noise. In low-noise conditions, convergence is rapid, and five iterations already provide a satisfactory estimate; as noise increases, convergence slows, and additional iterations continue to yield marginal but consistent improvements.

Beyond global fitting metrics, the iteration count has a more pronounced effect on the accuracy of modal identification under noisy conditions. For the numerical case with noise $= 2\%$, Modes~1--6 are consistently identified for both $N_\mathrm{iter} = 5$ and $N_\mathrm{iter} = 10$, with $f_n$ deviations below 0.06\%, $\zeta_n$ deviations below 2.40\%, and MAC values  $\geq 0.990$ in all cases, as reported in \cref{tab:results_iterations}. The two configurations are therefore essentially equivalent for the well-excited lower modes.

\begin{table}[htp!]
\centering
\small
\caption{Modal parameters identified by NExT-FRVF for the numerical case study with $noise = 2\%$ at two iteration counts. Below each value of $f_n$ and $\zeta_n$, the percentage difference with respect to the analytical solution is reported.}
\renewcommand{\arraystretch}{1.2}
\setlength{\tabcolsep}{5pt}
\begin{tabularx}{\textwidth}{l *{2}{>{\centering\arraybackslash}X}
                               *{2}{>{\centering\arraybackslash}X}
                               *{2}{>{\centering\arraybackslash}X}}
\toprule
& \multicolumn{2}{c}{\textbf{Natural Frequency} [Hz]}
& \multicolumn{2}{c}{\textbf{Damping Ratio} [--]}
& \multicolumn{2}{c}{\textbf{MAC} (diagonal values) [--]} \\
\midrule
\textbf{Mode} & $N_\mathrm{iter}=5$ & $N_\mathrm{iter}=10$
              & $N_\mathrm{iter}=5$ & $N_\mathrm{iter}=10$
              & $N_\mathrm{iter}=5$ & $N_\mathrm{iter}=10$ \\
\midrule
1 & 7.651  & 7.651  & 0.029 & 0.029 & 1.000 & 1.000 \\
  & (0.04\%) & (0.04\%) & (2.40\%) & (2.39\%) & & \\
2 & 47.952 & 47.952 & 0.030 & 0.030 & 1.000 & 1.000 \\
  & (0.05\%) & (0.05\%) & (0.37\%) & (0.36\%) & & \\
3 & 134.334 & 134.336 & 0.030 & 0.030 & 0.998 & 0.998 \\
  & (0.04\%) & (0.04\%) & (0.26\%) & (0.18\%) & & \\
4 & 263.681 & 263.680 & 0.030 & 0.030 & 0.998 & 0.998 \\
  & (0.05\%) & (0.05\%) & (0.04\%) & (0.08\%) & & \\
5 & 437.344 & 437.443 & 0.030 & 0.030 & 0.991 & 0.990 \\
  & (0.03\%) & (0.05\%) & (0.24\%) & (0.24\%) & & \\
6 & 657.523 & 657.260 & 0.030 & 0.030 & 0.992 & 0.993 \\
  & (0.06\%) & (0.02\%) & (0.18\%) & (0.08\%) & & \\
7 & 927.766 & 927.597 & 0.033 & 0.032 & 0.984 & 0.989 \\
  & (0.30\%) & (0.28\%) & (10.18\%) & (7.50\%) & & \\
8 & Not found & Not found & Not found & Not found & Not found & Not found \\
  & & & & & & \\
\bottomrule
\end{tabularx}
\label{tab:results_iterations}
\end{table}

For Mode~7, both iteration counts yield a successful identification, with natural frequency deviations of 0.30\% and 0.28\% and MAC values of 0.984 and 0.989 at $N_\mathrm{iter} = 5$ and $N_\mathrm{iter} = 10$, respectively. The principal difference between the two configurations lies in the damping ratio estimate for this mode: The error reduces from 10.18\% at $N_\mathrm{iter} = 5$ to 7.50\% at $N_\mathrm{iter} = 10$, indicating that additional pole relocations progressively refine the damping estimate for weakly excited modes even when detection itself is not compromised. Mode~8 remains unidentified in both configurations, consistent with the behaviour observed across the noise and weighting analyses, and attributable to noise contamination and limited observability in the available response set, rather than to insufficient iterations.

An additional consequence of increasing the iteration count is the reduction of spurious pole identifications, yielding a cleaner pole set and a more interpretable modal solution, even when the overall RMSE is already low. Taken together, these results indicate that RMSE convergence alone is not a sufficient criterion for assessing identification quality under noisy conditions: A higher iteration count can enhance both damping accuracy and pole set interpretability for low-energy modes, at the expense of a proportionally higher computational cost. In the experimental case study presented in \cref{sec:iss}, five iterations are retained as the baseline configuration, as they represent a practical compromise between complete identification and computational effort for noise levels encountered in ISS data.

\section{The International Space Station\label{sec:iss}}

\begin{figure}[htp!]
    \centering
    \includegraphics[width=0.7\linewidth]{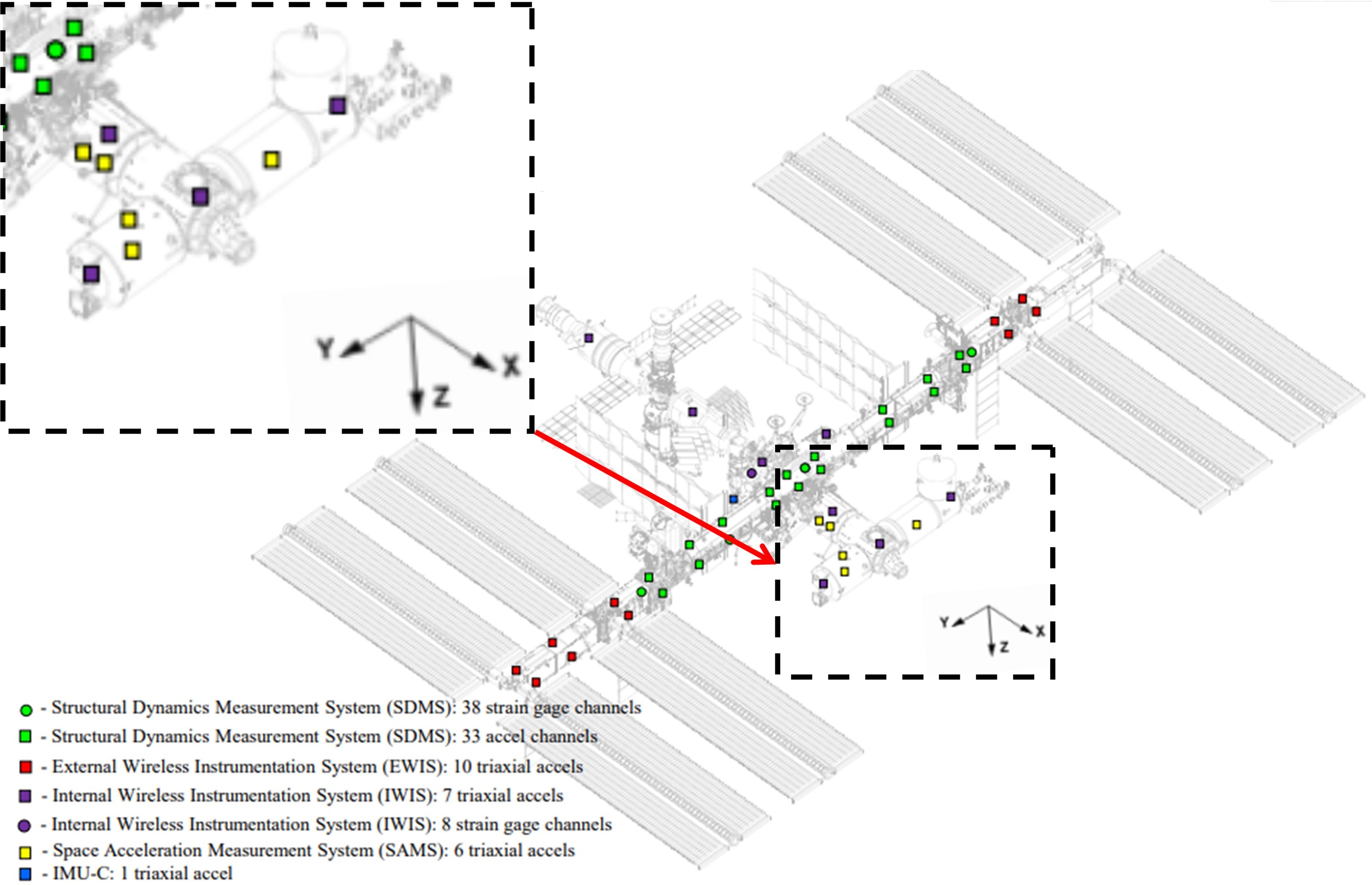}
    \caption{Sensor layout on the ISS. Yellow markers indicate the SAMS 
             sensor locations (adapted from \cite{Laible2018}).}
    \label{f6pos}
\end{figure}

The ISS provides a unique vibrational environment, approximating free-fall around Earth \cite{McPherson2015}, and constitutes an ideal case for output-only analysis, as moving the structure to a controlled laboratory setting for conventional modal tests would be practically impossible. Acceleration data are obtained using the Space Acceleration Measurement System (SAMS) \cite{GlennResearchCenter2024}, a network of triaxial accelerometers housed within isolated secondary structures to provide microgravity data and record small structural vibrations almost continuously, supporting the potential monitoring of station structural integrity. The NASA Glenn Research Centre (GRC) oversees these measurements. Since 2001, SAMS has been continuously recording excitation data, making it a key resource for both structural monitoring and microgravity research. All five SAMS triaxial sensors, namely known as 121f02-5 and 8, are considered in this study. Their layout and location on the ISS are shown in \cref{f6pos}.

Although the ISS centre of mass experiences near-zero gravity, its components are mechanically connected and vibrate individually. Accelerations arise from atmospheric drag, crew activities, or system operations, and propagate through structural joints or acoustically through the pressurised air. The ISS microgravity environment is characterised by three components: quasi-steady, vibratory, and transient regimes. This work focuses on the vibratory component, whose accelerations span the following frequency and magnitude range:
\begin{equation}
    0.01\,\text{Hz} \leq f \leq 300\,\text{Hz}, \qquad
    10\,\mu g \leq a \leq 1000\,\mu g
\end{equation}
The sources of these accelerations include rotating machinery (pumps, fans), crew routines (wake/sleep cycles, exercise), structural flexing, and turbulent airflow. Primary excitations also arise from vehicle structural modes, such as those of the truss and solar arrays, all of which contribute to the broadband vibratory regime of the station.

Mode identification focuses on natural frequencies between 0.01 and 2.5 Hz, which are expected to correspond to global structural modes of large flexible assemblies and to motion induced by crew activity. The available data correspond to a subset of the station instrumentation. Accordingly, the identified modes are of a local rather than global nature and are interpreted in that context throughout.

\begin{figure}[htp!]
    \centering
    \includegraphics[width=0.7\linewidth]{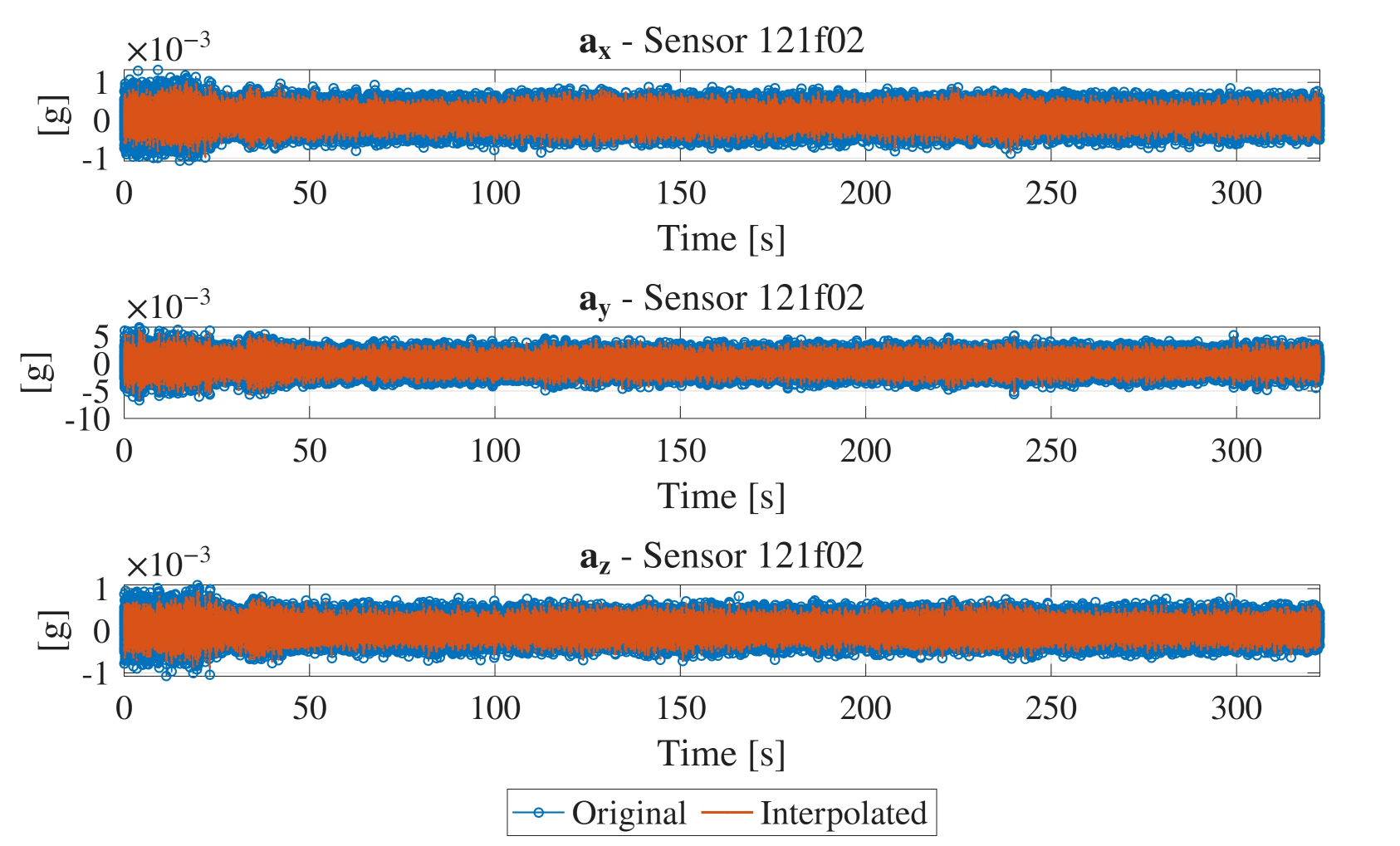}
    \caption{An example of original and interpolated acceleration signals: The three measurement channels of the triaxial sensor 121f02.}
    \label{f7accel}
\end{figure}

The dataset corresponds to the time series recorded on 18 August 2015, around 07:00 UTC. Not all sensors began recording simultaneously: The earliest, 121f03, starts at 
07:00:33.054, and the latest, 121f05, at 07:05:10.590. The analysis 
therefore uses the common recording interval from 07:05:10.590 to 
07:10:33.056. Within this window, the five triaxial ISS SAMS sensors share a sampling 
frequency $f_s = 500$ Hz and a duration $T_s = 5$ min 22.466 s, 
yielding a time resolution $\Delta t = 1/f_s = 0.002$ s and a frequency 
resolution $\Delta f = 1/T_s = 0.0031$ Hz, which is sufficient for the 
frequency range of interest. Although the common interval resolves the differing start times, samples from the various sensors within this interval are not synchronised to a common time base. This lag is corrected by interpolating each sensor record onto a shared uniform time vector at $f_s = 500$\,Hz via linear interpolation, with duplicate timestamps removed prior to resampling, as shown in \cref{f7accel}. The residual sub-sample timing misalignment introduced by this procedure is considered negligible relative to the frequency resolution $\Delta f = 0.0031$\,Hz of the analysis. This procedure is similar to that used by Boeing and NASA \cite{Laible2018}.

\begin{figure}[htp!]
    \centering
    \includegraphics[width=.7\textwidth]{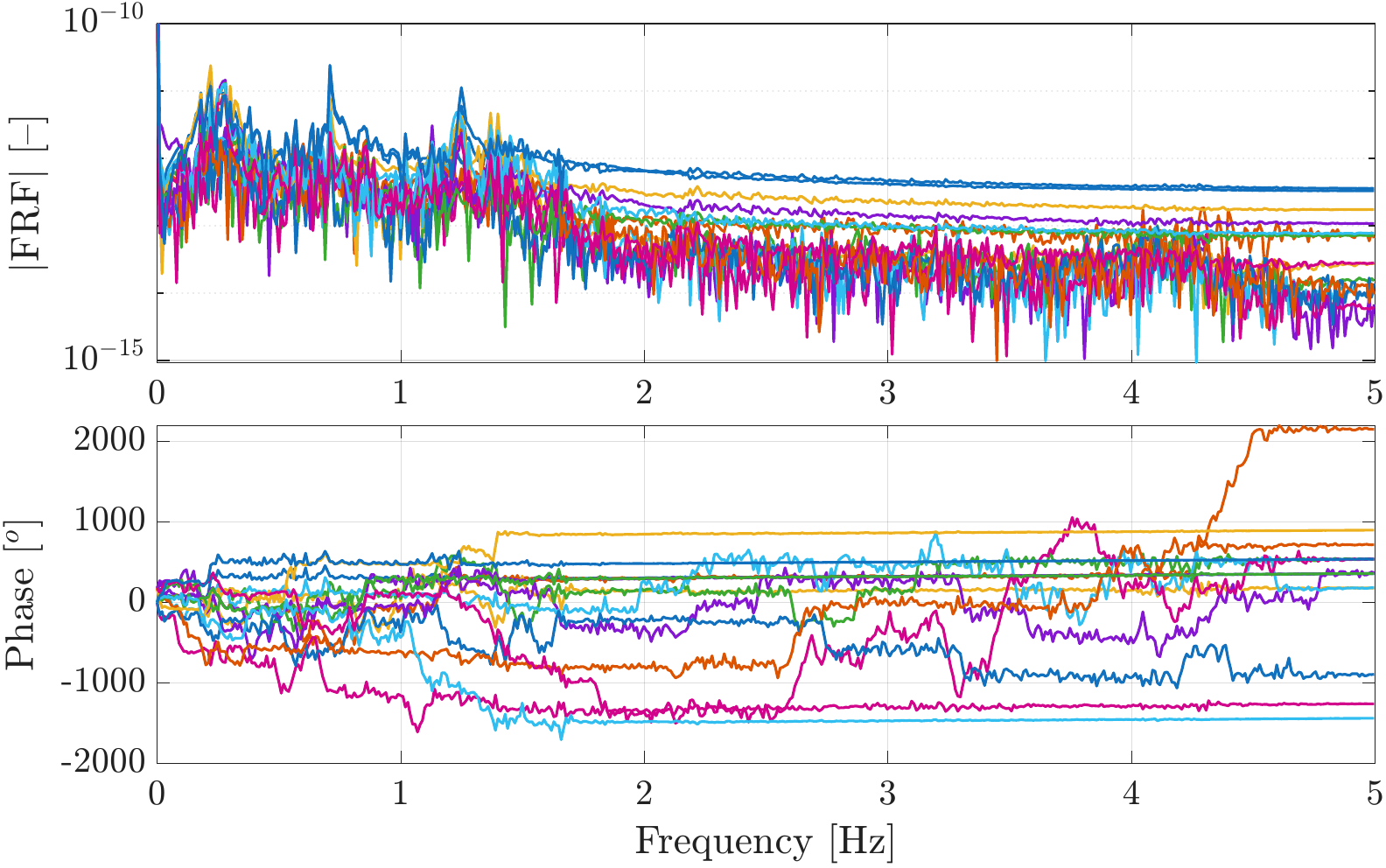}
    \caption{NExT-derived FRF from the fifteen output channels of the five triaxial ISS SAMS sensors.}
    \label{f8FRF}
\end{figure}

Finally, each sensor records accelerations in its own local coordinate 
frame. The Principal Investigator Microgravity Services (PIMS) 
\cite{McPherson2002} provides the rotation information required to map these measurements to the ISS global frame, using the standard aerospace 
yaw--pitch--roll direction cosine matrix:
\begin{equation}
\mathbf{R}_T = 
\begin{bmatrix}
    cYcP  & sYcP  & -sP  \\
    cYsPsR - sYcR & sYsPsR + cYcR & cPsR \\
    cYsPcR + sYsR & sYsPcR - cYsR & cPcR
\end{bmatrix}
\end{equation}
where $c$ and $s$ denote cosine and sine, respectively, and $Y$, $P$, $R$ are the angles of yaw, pitch, and roll, in the same order. After projecting all signals into this common reference frame, the preprocessed acceleration data are passed to the NExT-FRVF. With the preprocessed data, an SI process is performed, analogous to the numerical case (\cref{sec:nextfrvf}). The NExT algorithm is first applied to the acceleration data to obtain the FRF, shown in \cref{f8FRF}. 

Despite the irregular spectral background expected from broadband ambient excitation, several clearly distinguishable resonance peaks are visible. These can be subsequently compared with the $f_n$ of the identified modes to verify their correspondence. Before the NExT-derived FRFs are passed to the FRVF algorithm, the fitting parameters are set based on the parametric study conducted in the numerical case. The native sampling frequency of 500 Hz far exceeds what is required for modes up to 2.5 Hz; accordingly, the data are decimated to $f_s = 10$\,Hz using the MATLAB \texttt{decimate}\footnote{\url{https://uk.mathworks.com/help/signal/ref/decimate.html}} function, providing a margin above the Nyquist limit without introducing aliasing artefacts. For all three methods, the model order spans $N \in [20, 100]$ in steps of $\Delta k = 2$, selected to capture the expected modes shown in \cite{McPherson2015}, with the stabilisation criteria defined in \cref{sec:num}. Unit weighting is retained for NExT-FRVF: The frequency band of interest is narrow (0.01--2.5 Hz) and the SAMS acceleration amplitudes are uniformly low throughout the microgravity environment, so spectral amplitude variation within this band is minimal and inverse weighting provides no differential benefit. Five iterations are used, as convergence was observed consistently by the third iteration. The other two provide a conservative margin against noise-driven pole drift.

The FRVF fitting is then applied to the cross-spectral representation (NExT-derived FRF) of the acceleration signals. No analytical reference exists for the present output-only case. Although FEM and single-engine firing forced response results are available in \cite{Fitzpatrick2013}, a direct comparison is of limited validity given the fundamental difference in the excitation type (here, under ambient conditions and structural loading conditions). NExT-ERA and SSI (with Canonical Variate Analysis), both established output-only modal analysis methods, are therefore used as benchmarks for the proposed NExT-FRVF. NExT-ERA is configured with a Hankel matrix of $700$ rows and $800$ columns, a singular value truncation threshold of $150$, and a shift parameter of $10$. SSI is applied with a block Hankel matrix of $i = 14$ block rows, determined as $i = 2\lceil N_{\max}/n_{\mathrm{ch}} \rceil$, where $N_{\max} = 100$ and $n_{\mathrm{ch}} = 15$ is the number of output channels. The NExT correlation functions used by both NExT-based methods are estimated with channel~1 as the reference and a maximum lag of $1000$ samples, corresponding to $100$\,s at the decimated sampling frequency.
Three systematic challenges arise in the inter-method comparison: Different methods do not identify the same set of modes, so only common modes are retained for comparison. In certain instances, the same $\mathbf{\phi}_n$ is associated with two closely spaced frequencies, indicative of double identification. Moreover, NExT-ERA can occasionally produce modes with similar frequencies, but markedly different mode shapes $\mathbf{\phi}_n$, so temporal consistency, defined as repeated occurrence across successive datasets, is used as the primary screening criterion.

\begin{figure}[htp!]
    \centering
    \includegraphics[width=0.5\linewidth]{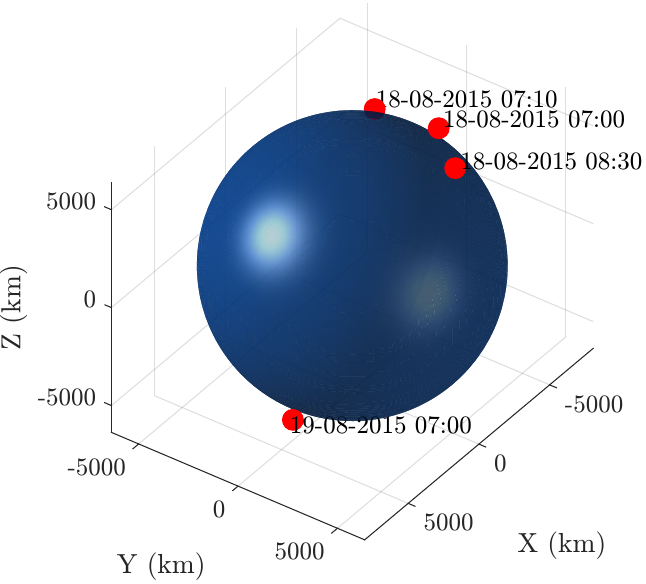}
    \caption{Location of the ISS at the four acquisition instants considered in this study.}
    \label{f10orb}
\end{figure}

To assess the robustness of the identification procedure, the consistency of the extracted modes is evaluated across four acquisition cases, selected to span a range of orbital and operational conditions:
\begin{itemize}
    \item \textit{Case~1}: Reference acquisition, 18 August 2015, 
    07:00:00 UTC;
    \item \textit{Case~2}: 10 minutes later, 18 August 2015, 
    07:10:00 UTC;
    \item \textit{Case~3}: 90 minutes later, 18 August 2015, 08:30:00 
    UTC, selected to account for possible thermally induced variations in 
    modal parameters over one orbital period;
    \item \textit{Case~4}: One day later, 19 August 2015, 07:00:00 UTC, 
    to assess the long-term repeatability.
\end{itemize}
The orbital positions of the ISS at these four acquisition instants are 
shown in \cref{f10orb}.

The NExT-FRVF results for the baseline case (Case~1) are validated against NExT-ERA and SSI in \cref{tab:modal_analysis}, which additionally includes reference frequencies from \cite{Fitzpatrick2013} -- obtained both via FEM and experimentally (Test) during the Stage~ULF4 S4-1A Dedicated Thruster Firing -- to contextualise the identified modes. Based on the mode labels and descriptions reported in \cite{Fitzpatrick2013}, the five identified modes correspond to the following physical deformation patterns: Mode~1 ($\approx$ 0.27 Hz) is a global station XY mode with Japanese Experiment Module (JEM) participation (7F1); Mode~2 ($\approx$ 0.30 Hz) is a station XZ mode (8F1); Mode~3 ($\approx 0.30$--$0.34$\,Hz) is also associated with station XZ motion (5F2), closely spaced with Mode~2 and likely reflecting the same global deformation shape under different excitation conditions; Mode~4 ($\approx 0.68$\,Hz) involves Russian Space Agency module XZ motion coupled with COL--JEM (COL -- Columbus module) outboard truss YZ deformation (8F2); and Mode~5 ($\approx 1.25$\,Hz) is a JEM XY and Node~3 XY mode with COL YZ participation (FEM mode~484). It is noted that the ISS configuration during the S4-1A DTF (October 2010) differs from that present during the ambient acquisition analysed here (August 2015). The reference frequencies are therefore used for mode labelling and physical interpretation only, and not as a quantitative validation target.

\begin{table}[htp!]
\centering
\small
\caption{Modal parameters identified by NExT-FRVF, NExT-ERA, and SSI          from the experimental case study (Case~1). Below each NExT-ERA and SSI value of $f_n$ and $\zeta_n$, the percentage deviation with respect to NExT-FRVF is reported. Reference values from \cite{Fitzpatrick2013}, obtained experimentally (Test) or via FEM.}
\setlength{\tabcolsep}{3pt}
\resizebox{\textwidth}{!}{  
\begin{tabular}{lcccccccccc}
\toprule
 & \multicolumn{4}{c}{\textbf{Natural Frequency} [Hz]} 
 & \multicolumn{4}{c}{\textbf{Damping Ratio} [--]} 
 & \multicolumn{2}{c}{\textbf{MAC} (matrix diagonal values) [--]} \\
\cmidrule(lr){2-5} \cmidrule(lr){6-9} \cmidrule(lr){10-11}
Mode & Ref. & NExT-FRVF & NExT-ERA & SSI 
     & Ref. & NExT-FRVF & NExT-ERA & SSI 
     & NExT-ERA & SSI \\
\midrule
1 & 0.259          & 0.272 & 0.265      & Not found 
  & 0.014           & 0.006 & 0.020      & Not found 
  & 0.649 & Not found \\
  & 7F1 (Test)     &       & (2.8\%)    & 
  & 7F1 (Test)     &       & (217.19\%) & & & \\
\addlinespace
2 & 0.308          & 0.297 & 0.299      & Not found 
  & 0.004          & 0.012 & 0.014      & Not found 
  & 0.997 & Not found \\
  & 8F1 (Test)     &       & (0.57\%)   & 
  & 8F1 (Test)     &       & (10.57\%)  & & & \\
\addlinespace
3 & 0.304          & 0.335 & 0.368      & Not found 
  & 0.013          & 0.007 & 0.021      & Not found 
  & 0.885 & Not found \\
  & 5F2 (Test)     &       & (9.88\%)   & 
  & 5F2 (Test)     &       & (181.08\%) & & & \\
\addlinespace
4 & 0.684          & 0.697 & 0.719      & 0.700 
  & 0.011           & 0.009 & 0.006      & 0.036 
  & 0.247 & 0 \\
  & 8F2 (Test)     &       & (3.1\%)    & (0.4\%) 
  & 8F2 (Test)     &       & (31.52\%)  & (294.57\%) & & \\
\addlinespace
5 & 1.251          & 1.242 & 1.242      & 1.237 
  & N.A.            & 0.009 & 0.004      & 0.017 
  & 0.849 & 0.043 \\
  & 484 (FEM)      &       & (0\%)      & (0.37\%) 
  & 484 (FEM)      &       & (50.56\%)  & (88.76\%) & & \\
\bottomrule
\end{tabular}}
\label{tab:modal_analysis}
\end{table}

The agreement $f_n$ between NExT-FRVF and NExT-ERA is strong across all five modes, with deviations not exceeding 9.88\%. SSI identifies only two modes, failing to detect any mode below 0.7 Hz, which limits its utility for global structural characterisation in this frequency range. Where SSI does identify a mode, frequency agreement is comparable to that of NExT-ERA, confirming that the frequency estimates are not method-dependent. $\zeta_n$ identification is unreliable across all methods, with NExT-ERA deviations relative to NExT-FRVF ranging from 10.57\% to 217.19\%, and SSI deviations reaching 294.57\%; these discrepancies are consistent with the known difficulty of output-only damping estimation under broadband, non-stationary excitation \cite{Fitzpatrick2013}, and preclude the use of damping as a validation metric in this context. $\mathbf{\phi}_n$ agreement, assessed via the MAC, is strong for Modes~2 and~3 (MAC\,$= 0.997$ and $0.885$, respectively, against NExT-ERA), satisfactory for Mode~5 (MAC\,$= 0.849$), and poor for Modes~1 and~4 (MAC\,$= 0.649$ and $0.247$), indicating that the latter two are either weakly excited, locally dominated, or subject to mode mixing given the limited spatial coverage of the available SAMS sensor configuration. SSI MAC values are consistently lower, dropping to zero for Mode~4; that highglights the method limited performance in this frequency range.

\begin{figure}[htp!]
    \centering
    \begin{subfigure}[b]{0.32\linewidth}
        \centering
        \includegraphics[width=\linewidth]{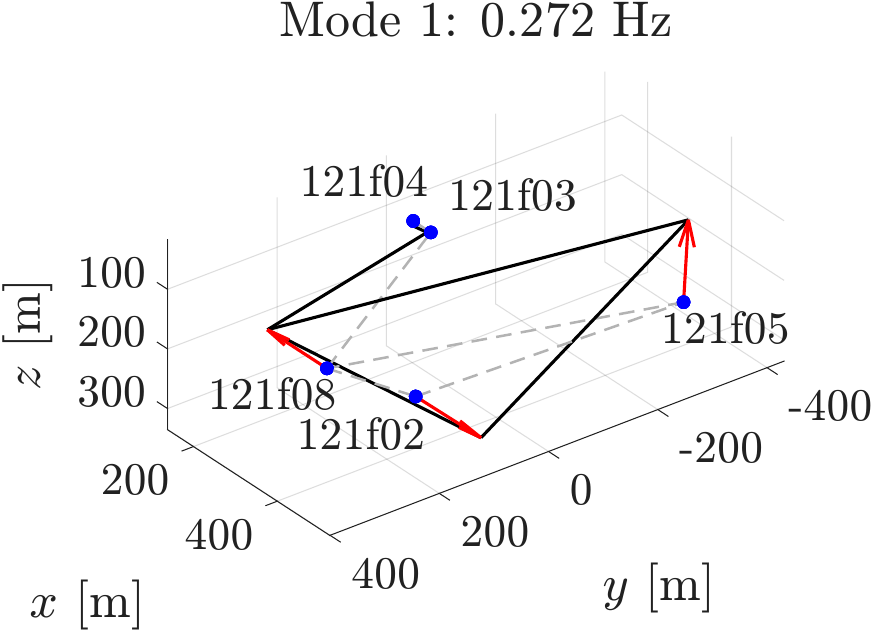}
        \caption{}
        \label{fig:mode1}
    \end{subfigure}
    \begin{subfigure}[b]{0.32\linewidth}
        \centering
        \includegraphics[width=\linewidth]{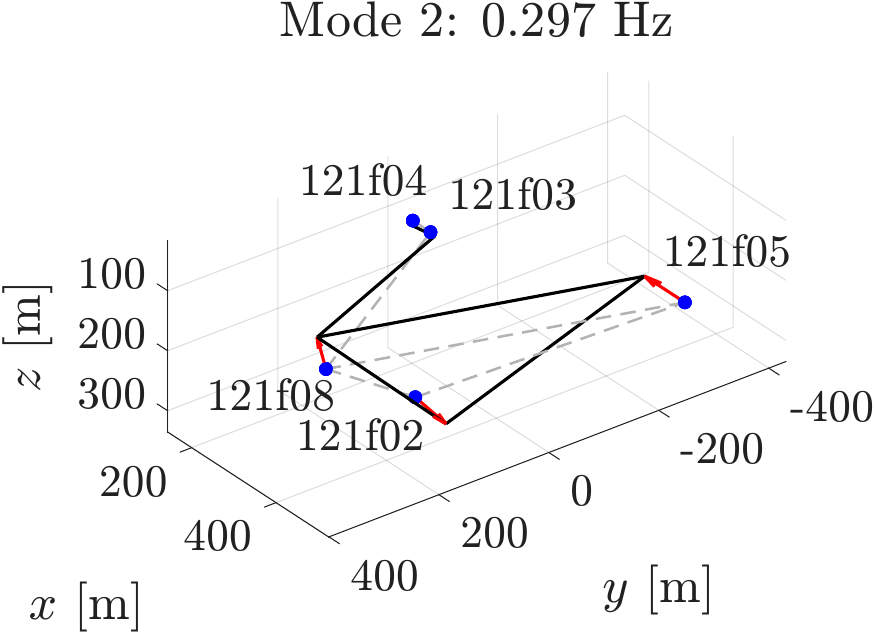}
        \caption{}
        \label{fig:mode2}
    \end{subfigure}
    \begin{subfigure}[b]{0.32\linewidth}
        \centering
        \includegraphics[width=\linewidth]{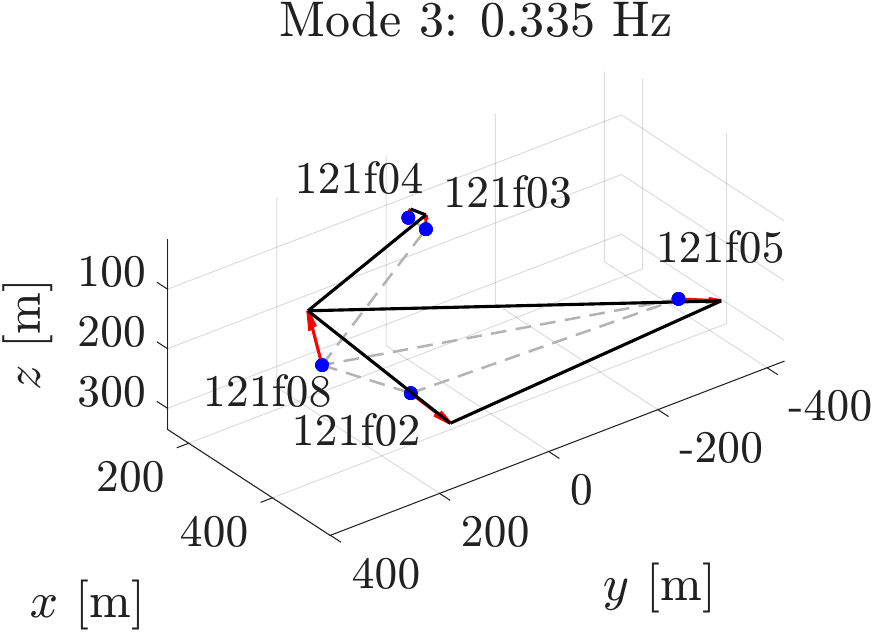}
        \caption{}
        \label{fig:mode3}
    \end{subfigure}
    \begin{subfigure}[b]{0.32\linewidth}
        \centering
        \includegraphics[width=\linewidth]{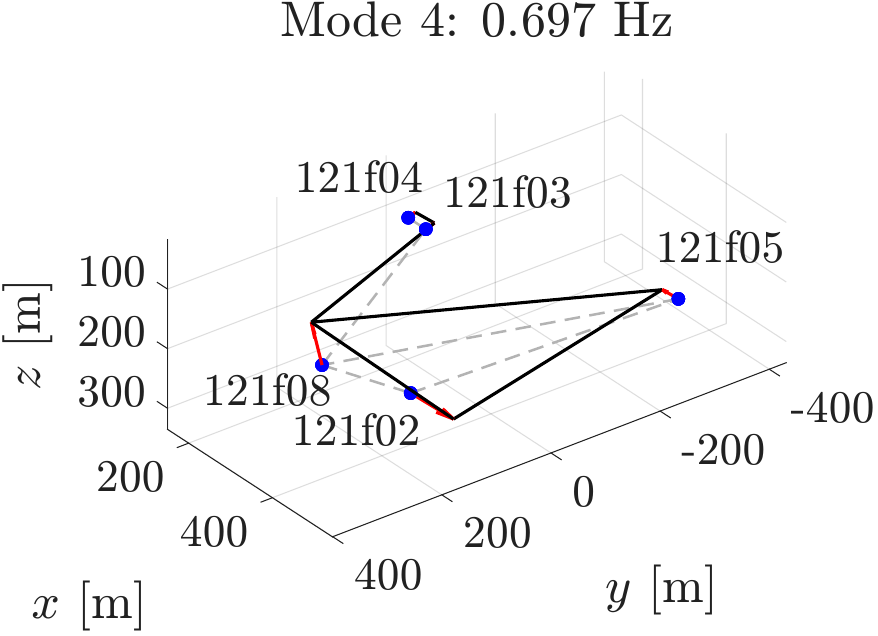}
        \caption{}
        \label{fig:mode4}
    \end{subfigure}
    \begin{subfigure}[b]{0.32\linewidth}
        \centering
        \includegraphics[width=\linewidth]{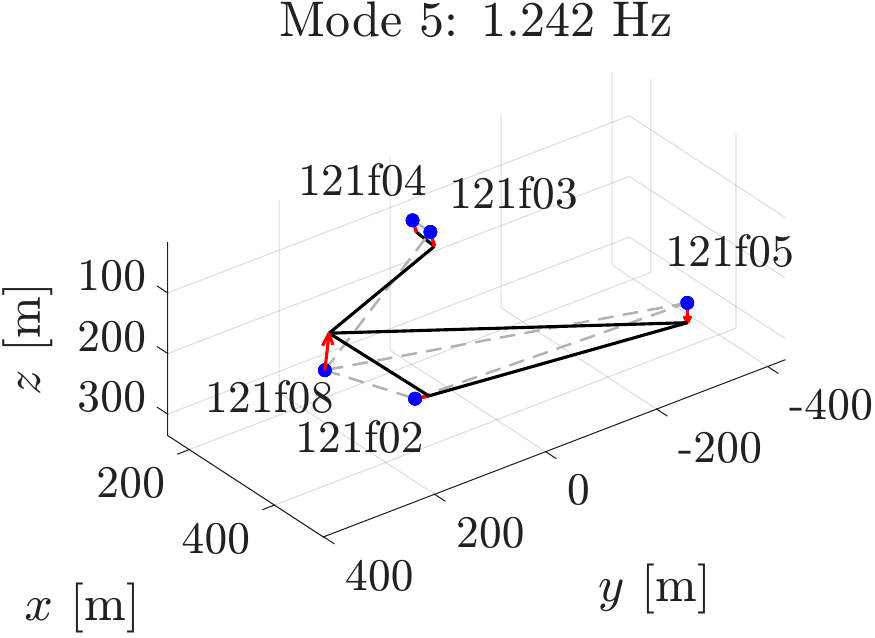}
        \caption{}
        \label{fig:mode5}
    \end{subfigure}
    \begin{subfigure}[b]{0.32\linewidth}
        \centering
        \raisebox{2.5cm}{\includegraphics[width=.5\linewidth]{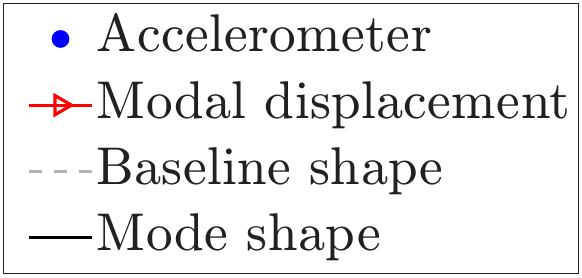}}
        \caption{}
        \label{fig:legend}
    \end{subfigure}

    \caption{ISS NExT-FRVF-identified mode shapes represented in three dimensions against the structural geometry: (a--e) first five mode shapes, and (f) associated plot legend.}
    \label{M5}
\end{figure}

The physical relevance of the identified modes is further assessed by examining the three-dimensional direction of motion reconstructed from the SAMS sensor measurements, mapped onto the actual structural geometry of the ISS and shown in \cref{M5}.  

The coupled bending-torsional behaviour observed in several modes is consistent with the structural constraints imposed by the surrounding ISS assembly, which enforces synchronised, lower-magnitude accelerations across the sensor cluster and supports the physical meaningfulness of the identified modes.

\begin{figure}[htp!]
    \centering
    \begin{subfigure}[b]{0.45\textwidth}
        \centering
        \includegraphics[width=\textwidth]{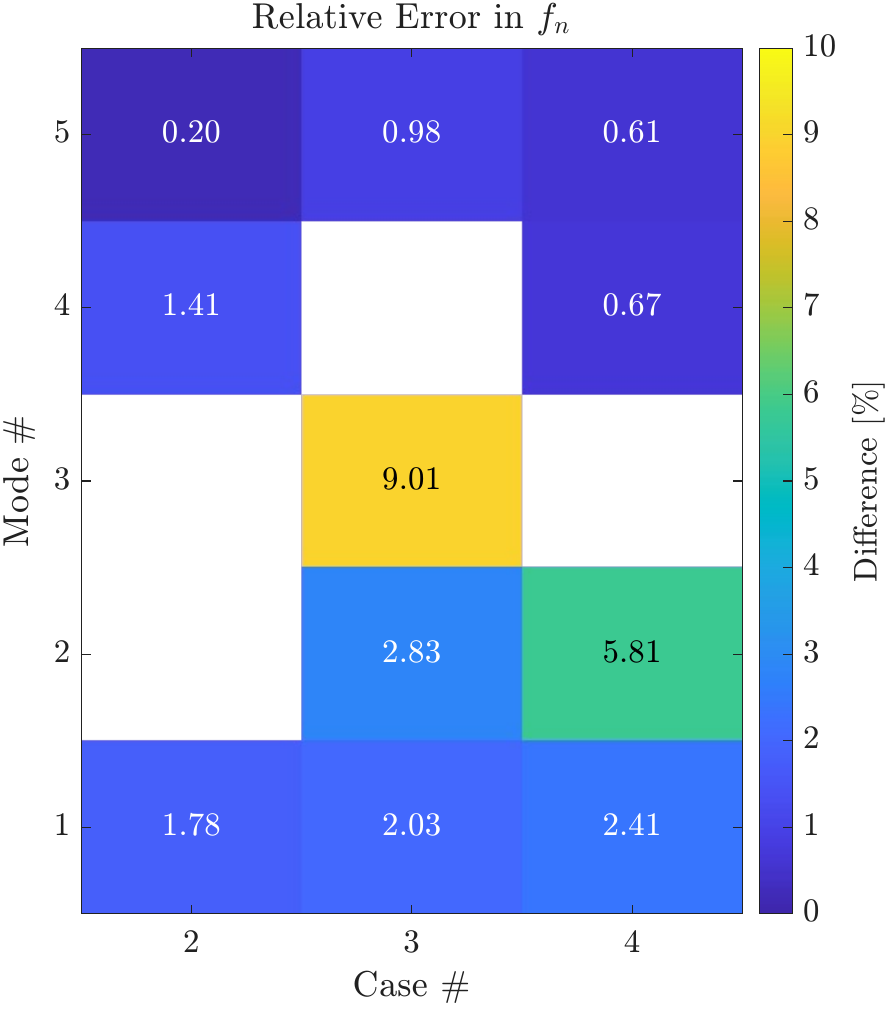}
        \caption{}
        \label{f91}
    \end{subfigure}
    \begin{subfigure}[b]{0.45\textwidth}
        \centering
        \includegraphics[width=\textwidth]{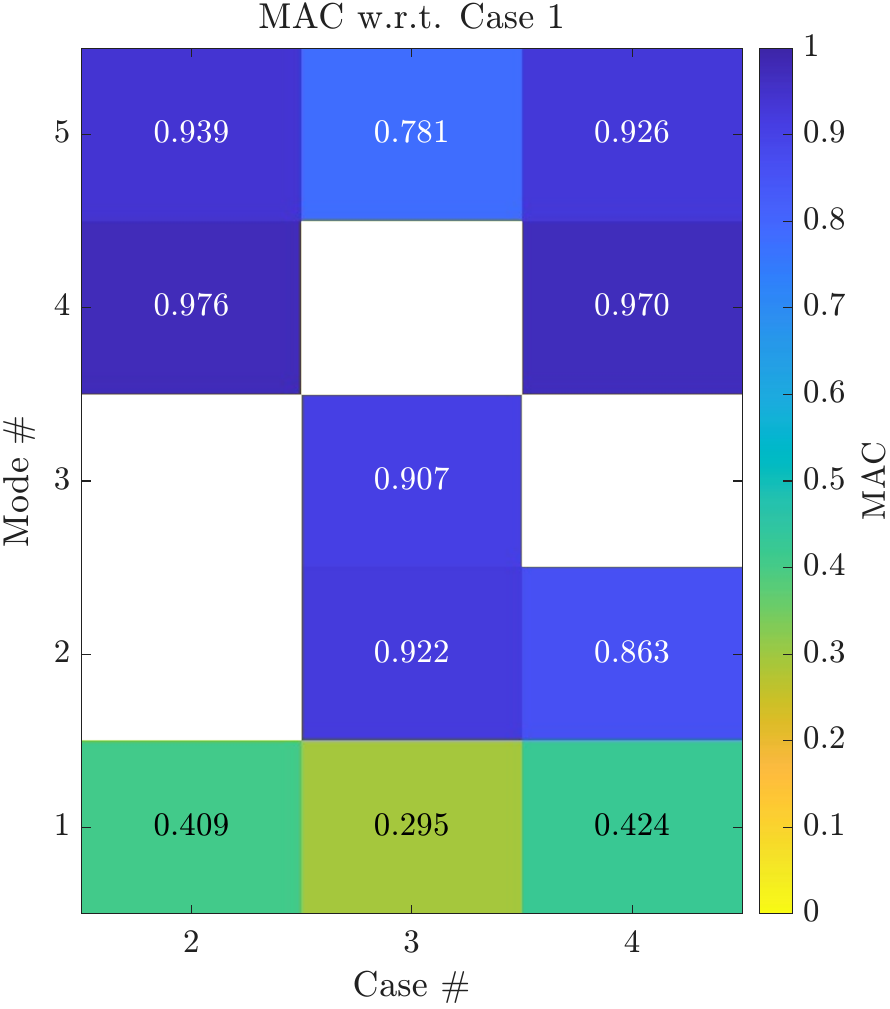}
        \caption{}
        \label{f92}
    \end{subfigure}
    \caption{NExT-FRVF modal parameter comparison between Cases~2, 3, and~4, with Case~1 as the reference: Natural frequency percentage deviation (a) and MAC (b). Blank entries represent unidentified or unmatched modes.}
    \label{fig:cases}
\end{figure}

After having assessed NExT-FRVF identification against benchmark methods in the baseline case, the temporal consistency of the extracted modes is evaluated across Cases~2--4 to demonstrate the modal tracking capability of the method. $f_n$ and MAC values relative errors to Case~1 are summarised in \cref{fig:cases}, where the blank entries represent unidentified or unmatched modes. Please note that due to the high variability of the identified $\zeta_n$, only $f_n$ and $\mathbf{\phi}_n$ are considered for this analysis. However, avoiding the use of damping ratios is aligned with the best practices and current recommendations in modal tracking (see, e.g., \cite{Pereira2020Role,Pereira2022ModalTracking}).

For the successfully matched entries of Modes~2--5, the MAC values are generally high, with most values above \(0.85\). The only lower matched value in this group is observed for Mode~5 in Case~3, where the MAC is \(0.781\), still relatively high for experimental data of this complexity.
{Hence, these Modes satisfy the NASA validity criteria of MAC$> 0.7$ and frequency deviation $< 10\%$ }\cite{Laible2018}.
On the contrary, Mode~1 exhibits consistently low cross-case MAC values (\(0.409\), \(0.295\), and \(0.424\) for Cases~2, 3, and~4, respectively), systematically below the NASA threshold of 0.7 across all follow-up cases. 
This behaviour is consistent with the poor mode shape agreement already observed for Mode~1 in the baseline case (MAC$= 0.649$ against NExT-ERA, \cref{tab:modal_analysis}), and reflects the same underlying factors: Limited spatial coverage of the available SAMS sensor configuration (which is evidently more impactful for this specific mode), local dominance of the mode shape, and potential mode mixing at this frequency. Possibly, this could also be due to intermittent noise contamination or sensitivity to crew activity and structural loading conditions at the time of acquisition.
Thus, the systematic failure of Mode~1 to meet the MAC threshold is more likely to reflect an inherent limitation of the available sensor subset, rather than an identification failure of the proposed method. Importantly, the estimated $\zeta_n$ are not used for validation purposes: Deviations across cases show no consistent pattern, reaching up to 60\%, consistent with the known sensitivity of output-only damping estimates to noise and non-stationarity, as acknowledged in NASA operational guidelines \cite{Laible2018}.

\section{Conclusions\label{sec:conc}}

Space structures exist in a very complex environment. Thus, ambient vibration testing is not as straightforward as in more conventional applications, such as civil structures and infrastructures.
Here, a novel output-only extension of the Fast and Relaxed Vector Fitting (FRVF) is proposed for this application. Pairing the FRVF with the Natural Excitation Technique (NExT) allows direct application to output acceleration recordings in the time domain.

The main findings of this work are as follows:

\begin{itemize}
    \item NExT-FRVF achieves strong agreement with NExT-ERA on a numerical Euler--Bernoulli beam, with natural frequency errors below 0.05\%;
    \item Damping errors remain below 1.35\% for all identified modes;
    \item For well-excited modes, the three weighting schemes yield     equivalent modal parameter estimates;
    \item Strong inverse weighting reduces the error in modal damping estimation and recovers higher modes, undetected under unit weighting;
    \item In narrowband, low-amplitude scenarios, all schemes would be equivalent. Thus, unit weighting remains the appropriate default;
    \item Five iterations are sufficient for reliable frequency and mode shape identification;
    \item Additional iterations yield marginal damping improvement for weakly excited modes;
    \item NExT-FRVF is validated on real International Space Station SAMS data, with strong mode shape agreement against benchmark methods;
    \item The modal parameters are consistent across the selected cases and comparable with the ranges reported in the existing literature.
    \item In particular, the identified mode shapes appear physically meaningful, considering the limited amount and spatial representativeness of the available channels, and aligned with the expected mode described in the literature.
    
\end{itemize}

Future work could focus on reducing computational cost, applying machine learning for parameter optimisation, developing automated clustering, and testing the method on damaged structures. Practically speaking, NExT-FRVF shows strong potential and real-life tested viability as an output-only System Identification method and strong potential for the structural health monitoring of aerospace structures in operation, including (but not limited to) spacecraft in orbit, and for supporting preventive maintenance for large and/or hard-to-access structures.

\appendix
\section{Euler--Bernoulli beam element matrices}\label{sec:appA}
\begin{equation}
\mathbf{M}_e = \frac{\rho A L}{420}
\begin{bmatrix}
156 & 22L & 54 & -13L \\
22L & 4L^2 & 13L & -3L^2 \\
54 & 13L & 156 & -22L \\
-13L & -3L^2 & -22L & 4L^2
\end{bmatrix}
\label{eq:mass}
\end{equation}

\begin{equation}
\mathbf{K}_e = \frac{1}{L}
\begin{bmatrix}
\frac{12 E I_y}{L^2} & \frac{6 E I_y}{L} & -\frac{12 E I_y}{L^2} & \frac{6 E I_y}{L} \\
\frac{6 E I_y}{L} & 4 E I_y & -\frac{6 E I_y}{L} & 2 E I_y \\
-\frac{12 E I_y}{L^2} & -\frac{6 E I_y}{L} & \frac{12 E I_y}{L^2} & -\frac{6 E I_y}{L} \\
\frac{6 E I_y}{L} & 2 E I_y  & -\frac{6 E I_y}{L} & 4 E I_y
\end{bmatrix}.
\label{eq:stiff}
\end{equation}

Note that the element matrices \(\mathbf{M}_e\) \ref{eq:mass} and \(\mathbf{K}_e\) \ref{eq:stiff} are defined consistently for an Euler--Bernoulli beam that undergoes transverse bending displacement along \(Z\), according to the reference frame portrayed in Figure \ref{fig:beam_main}, with the local DOFs ordered as

\[
\mathbf{q}_e =
\begin{bmatrix}
w_1 & \theta_1 & w_2 & \theta_2
\end{bmatrix}^{\mathrm{T}},
\]

where \(w_1\) and \(w_2\) are the transverse displacements at the two element nodes, and \(\theta_1\) and \(\theta_2\) are the corresponding cross-section rotations.

\section*{{\small Author Contributions}}
\noindent {\small {Conceptualisation, M.C. and G.D.; methodology, M.C.A., M.C., and G.D.; software, M.C.A., M.C., G.D., and S.U.E.; validation, M.C., and G.D.; formal analysis, M.C, G.D., M.C.A. and S.U.E.; investigation, M.C.A. and S.U.E.; resources, M.C., G.D., O.E.B.M.; data curation, M.C.A. and S.U.E.; writing---original draft preparation, M.C.A., S.U.E., M.C., and G.D.; writing---review and editing, M.C.A., M.C., G.D., and O.E.B.M.; visualisation, M.C.A., S.U.E., M.C., and G.D.; supervision, M.C. and G.D.; project administration, M.C. and G.D.; funding acquisition, M.C., G.D., O.E.B.M..}}

\section*{{\small Declaration of conflicting interests}}
\noindent {\small The author(s) declared no potential conflicts of interest with respect to the research, authorship, and/or publication of this article.}

\section*{{\small Funding}}
\noindent {\small
The first author is supported by the Centro Nazionale per la Mobilità Sostenibile (MOST – Sustainable Mobility Center), Spoke 7 (Cooperative Connected and Automated Mobility and Smart Infrastructures), Work Package 4 (Resilience of Networks, Structural Health Monitoring and Asset Management). 
The second author is supported by grant JDC2024-055593-I funded by MICIU/AEI /10.13039/501100011033 and by ESF+.
The authors from Universidad Carlos III de Madrid have been supported by the Madrid Government (\emph{Comunidad de Madrid} – Spain) under the Multiannual Agreement with the Universidad Carlos III de Madrid (\href{https://researchportal.uc3m.es/display/act564873}{IA\_aCTRl-CM-UC3M}).
The third author’s stay abroad during her graduate thesis was supported by the Erasmus+ Programme of the European Union under a student mobility grant.}

\section*{{\small Acknowledgements}}
\noindent {\small
The acceleration recordings from the International Space Station have been retrieved from \url{https://gipoc.grc.nasa.gov/wp/pims/home/} and are cited in this work. The authors thank the National Aeronautics and Space Administration (NASA) for making them openly available.

\section*{{\small Data Availability Statement}\label{sec:6_data}}
{\small 
The output-only implementation of Fast and Relaxed Vector Fitting (Data file 1) used in this work is openly available from the Zenodo repository at [\textit{DOI to be reserved after paper acceptance}], and the identification results and numerical dataset (Data file 2) supporting this study are openly available from the Zenodo repository at [\textit{DOI to be reserved after paper acceptance}]. In addition, this study used third-party (NASA) experimental data from the International Space Station (Data file 3) made available at \url{https://gipoc.grc.nasa.gov/wp/pims/home/} under a licence that the authors do not have permission to share.
This project contains the following underlying data:
\begin{itemize}
    \item Data file 1. A tutorial for Fast and Relaxed Vector Fitting operational modal analysis;
    \item Data file 2. Data supporting: International space station operational modal analysis via iterative pole relocation;
    \item Data file 3. NASA Principal Investigator Microgravity Services International Space Station acceleration recordings. 
\end{itemize}
Data files 1 and 2 are available under the terms of the [GNU General Public License v3.0 (GPL 3.0)].}

\bibliographystyle{elsarticle-num}

\end{document}